\documentclass[journal]{IEEEtran}

\ifCLASSINFOpdf
\else
\fi

\usepackage{cite}
\usepackage{tikz}
\usepackage{verbatim} 

\usepackage{nicefrac}
\usepackage{multirow}
\newtheorem{theorem}{Theorem}
\newtheorem{remark}{Remark}
\usepackage{dblfloatfix}

\usepackage{tikz}
\usetikzlibrary{positioning}
\usepackage{pgfplots}

\usepackage[multiple]{footmisc}
\usepackage{tikzscale}
\usepackage{xspace}
\usepgfplotslibrary{groupplots}
\pgfplotsset{compat=1.16}

\usepackage{hyperref}
\usepackage[switch]{lineno}
\usepackage{lipsum}
\newcommand{\vect}[1]{\mathbf{#1}}

\DeclareRobustCommand*\cal{\relax\mathcal}

\usepackage{caption}
\usepackage{subcaption}

\usepackage{amssymb}
\usepackage{amsmath}
\usepackage{flexisym}

\usetikzlibrary{external}
\usepackage{graphicx,xfp}
\usepackage{multicol}
\usepackage{multirow}
\usepackage{color}
\usepackage{xcolor}
\usepackage{booktabs}
\usepackage{graphicx}
\usepackage{nicefrac}
\usepackage{amsmath}
\usepackage{array}
\newcolumntype{P}[1]{>{\centering\arraybackslash}p{#1}}

\usepackage{algorithm2e}

\begin{document}
%
\title{Short-Term Load Forecasting Using AMI Data}
%
%
%






\author{Haris Mansoor,
        Sarwan Ali,
        Imdadullah Khan,
        Naveed Arshad,
        Muhammad Asad Khan,
        and Safiullah Faizullah
\thanks{ Haris Mansoor, Imdadullah Khan, and Naveed Arshad are with Department of Computer Science, Lahore University of Management Sciences (LUMS), Lahore, Pakistan, e-mail: \{haris.mansoor, imdad.khan, naveedarshad\}@lums.edu.pk}
\thanks{Sarwan Ali is with the Department
of Computer Science, Georgia State University, Atlanta, USA, e-mail: sali85@student.gsu.edu}
\thanks{Muhammad Asad Khan is with Department of Telecommunication, Hazara University, Mansehra, Pakistan, e-mail: asadkhan@hu.edu.pk}
\thanks{Safiullah Faizullah is with Department of Computer Science, Islamic University, Madinah, KSA, email: safi@iu.edu.sa}
\thanks{Manuscript received October 17, 2021; }}

%
%

\markboth{IEEE Internet of Things Journal,~Vol.~0, No.~0, February~2022}%
{Shell \MakeLowercase{\textit{et al.}}: Bare Demo of IEEEtran.cls for IEEE Journals}
%



\maketitle


\newpage

\begin{abstract}
Accurate short-term load forecasting is essential for the efficient operation of the power sector. Forecasting load at a fine granularity such as hourly loads of individual households is challenging due to higher volatility and inherent stochasticity. At the aggregate levels, such as monthly load at a grid, the uncertainties and fluctuations are averaged out; hence predicting load is more straightforward. This paper proposes a method called Forecasting using Matrix Factorization (\textsc{fmf}) for short-term load forecasting (\textsc{stlf}). \textsc{fmf} only utilizes historical data from consumers' smart meters to forecast future loads (does not use any non-calendar attributes, consumers' demographics or activity patterns information, etc.) and can be applied to any locality. A prominent feature of \textsc{fmf} is that it works at any level of user-specified granularity, both in the temporal (from a single hour to days) and spatial dimensions (a single household to groups of consumers). We empirically evaluate \textsc{fmf} on three benchmark datasets and demonstrate that it significantly outperforms the state-of-the-art methods in terms of load forecasting. The computational complexity of \textsc{fmf} is also substantially less than known methods for \textsc{stlf} such as long short-term memory neural networks, random forest, support vector machines, and regression trees.
\end{abstract}


%
\IEEEpeerreviewmaketitle

\section{Introduction}




Smart grid is a communication and control network on top of the electric grid to collect and analyze data from \textsc{i}o\textsc{t} devices at various grid elements. A fundamental task in power systems management is to accurately predict consumers' electricity demands based on past loads collected from smart meters. Inaccuracy in the demand estimate and the subsequent operational decisions may result in grid instability and sub-optimal resource utilization entailing high economic costs. {\em Long-Term Load Forecasting} (a few months to a year) is needed for power infrastructure planning~\cite{hong2016probabilistic}. However, operational decisions for smart grids have to be made within a short time and require {\em Short-Term Load Forecasting (\textsc{stlf})} (a few hours to days)~\cite{qiu2018ensemble}. Renewable energy resources are inherently intermittent in nature~\cite{wang2018lasso,wu2017data}, with their increasing integration in the energy mix, accurate \textsc{stlf} becomes even more crucial~\cite{malekizadeh2020short}.


Load forecasting can be categorized based on the spatial granularity of forecast, ranging from large scale (e.g., feeders or grids level) to fine-scale (e.g., individual consumer or household level). 
Forecasting at a smaller scale is more challenging due to the interplay of many factors~\cite{lusis2017short,malekizadeh2020short,amato2020forecasting,koprinska2015correlation}. Most of these factors are unknown or hard to measure, such as demographics, number, and daily schedules of residents in a household. It is well known that with increasing spatial aggregation, the forecasting error decreases~\cite{kong2019short,amato2020forecasting}. This behavior is because many fluctuations tend to average out at larger scales.

A wide variety of methods have been proposed for \textsc{stlf} at large scales. These methods apply statistical and machine learning techniques to historical load data to predict the (total) load of all consumers~\cite{hong2016probabilistic,shi2018deep,JIANG2017hybrid,AMJADY2009short}. However, efficient and accurate load forecasting for short-term and at a fine scale is pivotal for demand response programs~\cite{hong2016probabilistic,kong2019short,qiu2018ensemble}, peak shaving~\cite{shi2018deep,haben2014new}, dynamic pricing~\cite{de2011short} and soft load-shedding schemes~\cite{tayyab2018smartcities,sarwan2019soft}.


In the {\em Advanced Metering Infrastructure} (\textsc{ami}), consumers are increasingly connected to the grid through smart meters, a typical example of \textsc{i}o\textsc{t}. Over 100 million smart meters have been installed in the USA~\cite{smart_meter_2022}. \textsc{ami} makes available short duration load data for individual consumers. There is an increasing research interest in utilizing this data for \textsc{stlf} to optimize power system management~\cite{haben2014new,koprinska2015correlation}.



\begin{figure}[ht!]
	\centering
	\includegraphics[page=1, width=\linewidth]{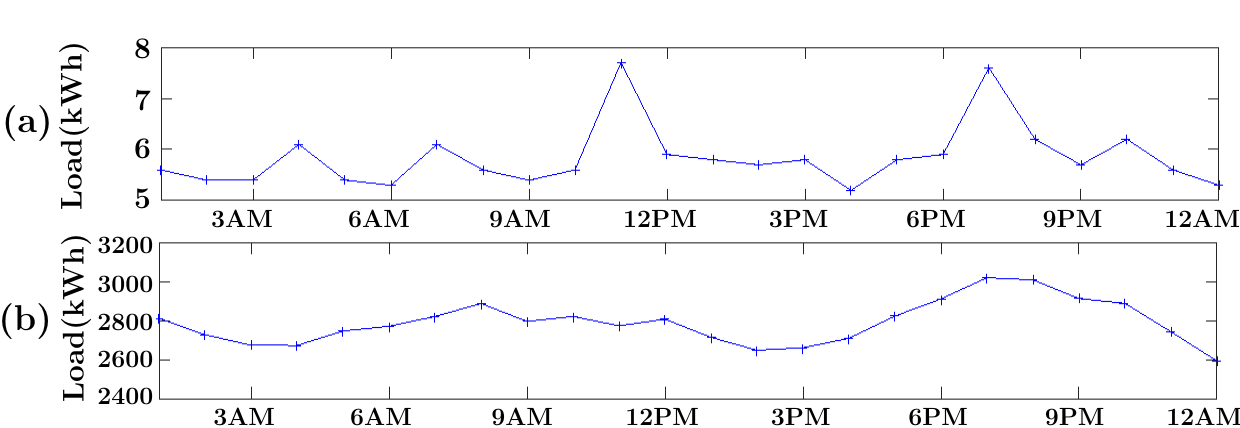}
	\caption{(a) Hourly load of a randomly selected household in Sweden dataset for a single day (Fri, $5$ Jan, $2004$). (b) Total hourly load of all households for the same day.}
	\label{sweden_data_analytics_with_aggregation}
\end{figure}

Most recent works on \textsc{stlf} group consumers into clusters based on their load profiles (hourly \textsc{ami} readings) and sociological data (obtained from surveys). Then the load of each cluster is predicted using machine learning methods~\cite{gajowniczek2017electricity,hsiao2015household,kell2018segmenting,QuilumbaLHWS15}. The survey data is not necessarily reliable when available and limits the applicability of the methods to a specific locality. Moreover, the total load of a cluster is smoother and less volatile (Figure~\ref{sweden_data_analytics_with_aggregation}), hence relatively easier to forecast.

In this paper, we propose an algorithm called {\em Forecasting using Matrix Factorization (\textsc{fmf})} for short-term load forecasting at any user-specified level of spatial and temporal granularity. We perform a comprehensive exploratory analysis of three benchmark datasets from three countries to visually observe patterns in the data. \textsc{fmf} first applies data transformation on the past load values to reduce skewness in the data. Since the dimensionality of data is high, to avoid the "curse of dimensionality", we factorize the load matrix using Singular Value Decomposition (\textsc{svd}) to get a low-dimensional representation both for time-stamps (hour) and consumers. The new feature vectors for time-stamps are based on all consumers' overall behavior in that hour. Similarly, the feature vectors of consumers are determined by their consumption patterns for the total period of the training data. (Note that we use the word consumer and household interchangeably throughout the paper). We empirically demonstrate that this representation improves the forecasting performance of our model. In this feature space, we aggregate hours into demonstrably more meaningful clusters. Finally, to predict the load of a consumer in a query hour (in the test set), we find clusters of hours in the training data that are {\em `most similar'} to the query hour and report an {\em `average'} of the consumer's loads in those clusters. 

We perform an extensive empirical evaluation of \textsc{fmf} to showcase its efficacy as a general-purpose forecasting method that works for a wide range of user-specified spatial and temporal granularity. We compare \textsc{fmf} with baseline and state-of-the-art (\textsc{sota}) methods using their respective experimental setup and evaluation metric. We show that \textsc{fmf} significantly outperforms known methods on \textsc{stlf} at the individual consumer level. We test \textsc{fmf} on forecasting loads for longer durations (up to a day), for groups (clusters) of consumers, and both combined. Results reveal that the performance of \textsc{fmf} surpasses the computationally expensive specialized methods for these tasks.

Key features of \textsc{fmf} are the following:

\begin{itemize}
	\item \textsc{fmf} only utilizes the readily available hourly electricity consumption \textsc{ami} data and does not require any consumers specific information. Hence \textsc{fmf} is more generally applicable. 
	\item \textsc{fmf} works at any level of user-specified granularity, both in the temporal and spatial dimensions.
	\item \textsc{fmf} does not require any complex training of the model, unlike Long Short Term Memory (\textsc{lstm}), Random Forest (\textsc{rf}), Support Vector Machine (\textsc{svm}), and Regression Tree (\textsc{rt}). Thus \textsc{fmf} is more suitable for \textsc{stlf}, where timely decisions need to be made.
	\item \textsc{fmf} achieves up to $26.5\%$ and $24.4 \%$ improvement in Root Mean Square Error (\textsc{rmse}) over \textsc{svr}  and \textsc{rt},   respectively and up to $73.8 \%$ and $38 \%$ improvement in Mean Absolute Percentage Error (\textsc{mape}) over \textsc{rf} and \textsc{lstm}, respectively.
	\item The representations learned for consumers and time-steps in our model preserves the overall structure of data, reduce the computational complexity of the model, and make the model scalable to big data. 
\end{itemize}

The rest of the paper is organized as follows. We provide a brief review of existing methods for \textsc{stlf} in Section~\ref{relatedwork}. In Section~\ref{Proposed_methodology} we describe the \textsc{fmf} scheme. Section~\ref{Experiments} contains datasets description and experimental setup. Results of empirical evaluation and comparisons of \textsc{fmf} are reported in Section~\ref{results_and_comparisons}. Finally, we conclude the paper in Section~\ref{Section_Conclusion}.

\section{Related Work}\label{relatedwork}
The existing work on \textsc{stlf} can be divided into three categories, namely (i) \textsc{stlf} at system or subsystem level in which aggregated the load of all consumers at a locality is forecasted, (ii) Clusters level \textsc{stlf}, where consumers are intelligently grouped into clusters and loads for those clusters are forecasted, and (iii) \textsc{stlf} for individual consumers in which load is forecasted separately for each customer.

\subsection{\textsc{stlf} at System or Subsystem Level}
Short-term load forecasting at a system/subsystem level is well explored in the literature. A neural network based method for \textsc{stlf} is proposed in~\cite{li2015short} in which households are grouped based on location, nature, and size of loads. In~\cite{zhang2016composite} $k$-nearest neighbors based algorithm is used to forecast day-ahead loads of groups of consumers. A framework using wavelet transform and Bayesian neural network for \textsc{stlf} at the system level is proposed in~\cite{ghofrani2015hybrid}. A time-series method using intra-day and intra-week seasonal cycles is proposed in~\cite{taylor2008evaluation} for forecasting country loads a few minutes ahead. In~\cite{chakhchoukh2011electric}, stochastic properties of electricity in France are utilized to predict short-term aggregated load. A kernel-based support vector regression model for the \textsc{stlf} at system level is proposed in~\cite{che2014short}. 
A hybrid machine learning model is proposed in~\cite{saxena2019hybrid}, which uses the combination of \textsc{arima}, logistic regression, and artificial neural networks to forecast day-ahead peak electric load at the system level.
Several authors proposed hybrid methods involving data preprocessing with classification, regression, and other machine learning based methods for \textsc{stlf} at system/subsystem level~\cite{li2016ensemble,malekizadeh2020short}.

There are several problems with directly using machine learning models for \textsc{stlf}, such as difficulties in parameter selection and non-obvious selection of input variables~\cite{tayab2020short}. Therefore, these models have to be combined with statistical models and different data preprocessing techniques to reduce the computational overhead. Authors in~\cite{tayab2020short} use a combination of neural network and statistical models for \textsc{stlf} at microgrid level. An unsupervised machine learning model is proposed in~\cite{yang2019short}, which combines Auto Correlation Function and Least Squares Support Vector Machines model to forecast short term load at the system level. The actual deployability of the forecasting algorithms at the country level is studied in~\cite{liao2020multiple}. The authors focus on prediction power, the robustness of model, the dependence of model on the dataset, and storage size. Multiple wavelet convolution neural networks are proposed, which balance these measures. Since the impact of variables on the demand changes over time, an online continuous learning approach is proposed and tested by~\cite{zamee2021online}. The author used correlation analysis to figure out the effect of variables and then used a neural network for prediction in an online setting.

\subsection{\textsc{stlf} for Clusters of Consumers}
A wide variety of methods utilize the increasingly available \textsc{ami} data to intelligently group the households into clusters (based on their consumption patterns) and forecast the load of these clusters. For cluster loads prediction, machine learning models such as random forest, neural networks, and deep learning are commonly used. Clustering is accomplished based on  similarities in load profiles (consumers' \textsc{ami} readings)~\cite{QuilumbaLHWS15} and consumers demographic information~\cite{kell2018segmenting}. Practice theory of human behavior is incorporated for improved clustering~\cite{StephenTHGJ17}, resulting in an accuracy boost for day-ahead system-level load forecast. A deep neural network-based model for \textsc{stlf} at an individual and subsystem level is proposed in~\cite{ryu2017deep}, which learns complex relations between weather, calendar, and previous consumption for individual households. A hybrid approach consisting of a convolutional neural network and $k$-means clustering algorithm is proposed in~\cite{dong2017short} to forecast the hourly load of clusters of households. In~\cite{li2016multi} the authors propose a multi-resolution clustering method to forecast half hourly load for households. The relationship between cluster size and forecast accuracy is studied in~\cite{da2013impact} using two forecasting methods, namely Holt-Winters and Seasonal Naive.

\subsection{\textsc{stlf} for Individual Consumers}
\textsc{stlf} at individual consumers' level is significantly more challenging due to high volatility and variability in load profiles~\cite{gajowniczek2014short}. The classical methods treat each consumers' data as a stationary time series for prediction. Time series methods for \textsc{stlf} use Kalman filter~\cite{ghofrani2011smart}, advanced statistical techniques~\cite{athukorala2010estimating}, and the standard Auto Regressive Integrated Moving Average (\textsc{arima}) forecasting models~\cite{shi2018deep}. These time series approaches, however, do not capture the complex nonlinear relationship between electricity consumption and periodic routines of household residents~\cite{hsiao2015household}. It has been shown in~\cite{veit2014household} that time series based methods hardly beat persistent forecast (using previous hour load value as the forecast for the next hour). Authors in~\cite{ignatiadis2019forecasting} propose a regression based method to forecast the monthly aggregated load of individual households. However, they do not take into account the temporal order of the historical loads in which the load values were observed. This limitation restricts the application of the method in any realistic scenario. 
A Pooling based deep recurrent neural network method is proposed in~\cite{shi2018deep} to predict individual households loads and improve upon the accuracy of \textsc{arima}, and other machine learning models. Similarly, \cite{kong2019short} uses \textsc{lstm} network along with density based clustering for household load forecasting. Authors in~\cite{mocanu2016deep} used a factored conditional restricted Boltzmann machine for load forecasting of buildings and showed improvement over the support vector machine and neural network.
A \textsc{stlf} model for individual household level is proposed in~\cite{gajowniczek2014short}, which uses standard machine learning models such as neural networks and \textsc{svm} for load forecasting.
\cite{yildiz2018household} proposes a model for \textsc{stlf}, which uses historical loads and weather information along with the information contained in typical daily consumption profiles (loads in mornings, evenings, and nights, etc.) for load forecasting.
A predicted model (sparse high-dimensional partially linear additive models) for \textsc{stlf} at individual households level is used in~\cite{amato2020forecasting}, which forecasts half-hourly electricity load for one day ahead.
In~\cite{chaouch2013clustering}, the authors propose a clustering based method, which uses historical load to forecast day ahead loads of individual households.
Hybrid methods for household level \textsc{stlf} incorporate additional activities patterns information (survey, demographic information etc.) to improve household load forecast~\cite{singh2012load,gajowniczek2017electricity}. However, the activity patterns information is not readily available in many cases.

\section{Proposed Approach} \label{Proposed_methodology}
In this section, we describe the detailed algorithm of \textsc{fmf}. \textsc{fmf} takes the load matrix $X \in \mathbb{R}^{m \times n}$ as input, with rows and columns corresponding to $m$ hours and $n$ consumers, respectively. The entry $X(i,j)$ is the electricity consumption of consumer $j$ at hour $i$. \textsc{fmf} broadly performs the following steps: We first preprocess the data to improve its statistical properties. Then we split $X$ into two submatrices $A$ and $B$. The submatrix $A$ consists of the first $m_1$ rows and is used as training data. While submatrix $B$, consisting of the last $m_{2} = m - m_1$ rows, is used for testing. Thus the dimensions of $A$ and $B$ are $m_1\times n$ and $m_2 \times n$, respectively. Next, we map hours of $A$ into a low-dimensional feature space based on the overall consumption patterns. In this feature space, hours of $A$ are clustered. The forecast for $B(i,j)$ is an {\em `average'} of consumer $j$'s loads (along with the loads consumers similar to $j$) in the $t$ clusters of hours of $A$ that are the most {\em `similar'} to the query hour $i$. However, since clusters are in a load-based (abstract) feature space and hour $i$ is in the testing period, we cannot use load values at hour $i$. Therefore, we find a common representation both for testing hours and clusters of training hours based only on the calendar attributes of hours. Similarities between query hours and clusters are evaluated with this representation. We provide details of each step below.

\subsection{Data Preprocessing}
\begin{figure}[ht!]
	\centering
	\includegraphics[page=2,width=\linewidth]{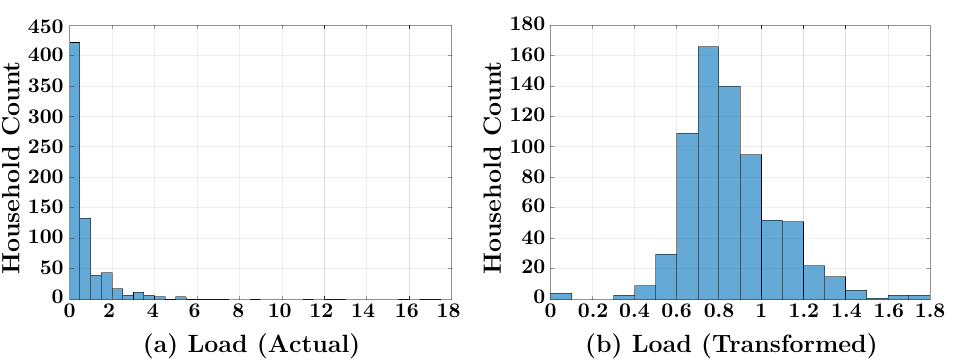}
	\caption{Consumers load distribution (Ireland dataset) for a randomly chosen hour (a) before and (b) after transformation.}
	\label{Irish_Histograms_All_Customers}
\end{figure}

First, we min-max normalize columns of $A$ to make the values unitless and scale them to $[0,1]$. The load values in $X$ are for individual consumers and a short duration; most are very low values. Hence there is a significant right skew and variation in the data. We apply the standard $q^{th}$ root transformation on $X$ as a preprocessing step i.e. every value $X(i,j)$ is replaced with $X(i,j)^{\nicefrac{1}{q}}$. For notational convenience, we still denote the transformed matrix by $X$. The skew and effect of transformation are depicted in Figure~\ref{Irish_Histograms_All_Customers}, showing load distributions at a randomly chosen hour. It is clear from the figure that a large number of values are very close to $0$ before transformation and that the transformed data is more {\em `normally'} distributed. The optimum value of $q$ is selected empirically as $4$, $5$, and $3$ for Sweden, Ireland, and Australia datasets, respectively.

\begin{remark}
Note that we apply the corresponding reverse transformation ($q^{th}$ power) after forecasting the load and report our predictions and their errors in the original scale.
\end{remark}

\subsection{Consumption Patterns based Feature Map and Clustering of Training Hours} \label{section_Matrix_Factorization}
Our goal is to cluster hours of $A$ based on the overall consumption patterns during these hours. Moreover, we want to define a similarity measure between a testing hour and a cluster of training hours. However, every hour of $A$ is (potentially) a very high dimensional vector ($n$). In such a high dimensional space, due to {\em curse of dimensionality}, no notion of pairwise similarities and thus clustering is significant. Forecasting accuracy, however, critically depends on the quality of clustering. 
To deal with this problem, we reduce dimensionality of the matrix $A$,
 using the following fundamental result on singular value decomposition (\textsc{svd}) from linear algebra [c.f.~\cite{strang1988linear}].
\begin{theorem}\label{svdTheorem} Suppose $Z$ is an $a\times b$ real matrix with rank $r$. Then there exists a factorization $Z = U\Sigma V^T$  such that $U$ is  ($a\times r$) matrix of orthonormal rows, $\Sigma$ is  $r \times r$ diagonal matrix of non-negative real numbers, and $V$ is  $b\times r$ matrix of orthonormal rows.
\end{theorem}


The rows of $U\Sigma$ and columns of $\Sigma V^T$ represent rows and columns of $Z$, respectively, in an abstract feature space (latent factors on which the data varies). The relevance of features is quantified by the singular values $(\Sigma)$. For $d\leq r$, let $\Sigma_d$ be the $d\times d$ diagonal submatrix of $\Sigma$ consisting of the largest singular values (these singular values contain maximum weight or energy in $\Sigma$~\cite{strang1988linear}). Let $U_d$ and $V_d$ be the submatrices of $U$ and $V$ consisting of the $d$ columns corresponding to values in $\Sigma_d$. The truncated matrix $Z_d =U_d\Sigma_d{V_d}^T$ approaches $Z$ as $d$ approaches $r$ and is called {\em low rank approximation} of $Z$.


We apply \textsc{svd} on $A$ to get $A = U\Sigma V^T$ and consider its approximation $A_d := U_d \Sigma_d {V_d}^T$. Let $H := U_d \Sigma_d$, then rows of $H$ are the $d$-dimensional representations of hours of $A$. Each column of $H$ is a feature of hours, based on the loads of all consumers in $A$. We cluster the training hours of $H$ into $r$ clusters, ${\mathcal C} = \{C_1,C_2,\ldots,C_r\}$. Each cluster $C_i$ contains hours that are substantially similar to each other based on the overall consumption patterns (see Figure~\ref{fig_clust_goodness_swe} in appendix).  

We choose the values of $d$ (the number of reduced dimensions) and $r$ (the number of clusters) by evaluating the quality of clusters. The chosen values for $r$ are $80$, $70$, and $70$, while those for $d$ are $300$, $424$, and $34$ for Sweden, Ireland, and Australia datasets respectively.
\begin{remark}
No dimensionality reduction is applied for the Australia dataset as the number of consumers (columns of the load matrix) is already small, i.e. "$34$".
\end{remark}

\begin{remark}
We selected the singular values $d$ of $\Sigma$ for \textsc{svd} that preserve $80\%$, $80\%$, and $100\%$ cumulative energy in $\Sigma$ for Sweden, Ireland, and Australia datasets, respectively. We can see in Figure~\ref{fig_svd_val_swe} (in appendix) that $d = 300$ (for $d \leq r $) contains $80\%$ of the cumulative energy of $\Sigma$ for Sweden dataset (Figure~\ref{fig_svd_val_swe} (b) in appendix).
\end{remark} 



We use $k$-means++ algorithm~\cite{arthur2007k} to cluster the rows of $H$. To avoid local minima, we replicate the clustering $1000$ times and select the most accurate partition.

Given a customer $j$, we now find other customers similar to $j$.
For this, we first take the transpose of matrix $X$ (to represent customers in rows and hours in columns) and apply \textsc{svd} to reduce the number of columns (hours). However, we perform this task for each month's hours separately (to capture the seasonality information).
To do this, we separate the hours of each month ($\approx 720$ hours for each month), which will give us a separate $\approx n\times 720$ dimensional matrix. On each month's matrix, we apply \textsc{svd} separately columns (and get $10$ principal components) to reduce its dimensions (we get $n\times 10$ dimensional matrix). 
Then we concatenate values of all months (all $n\times 10$ matrices together for 12 months) to form a single $n\times 120$ dimensional matrix (where $10 \times 12 = 120$). We refer to this as the seasonal \textsc{svd} approach. The top $k$ similar consumers (rows in this new matrix) to consumer $j$ are then selected using the Euclidean similarity. We use $k = 3$, empirically set using standard validation set approach~\cite{validationSetApproach}).

\subsection{Calendar Attributes Based Feature Map}
We use rows of $H$ to represent corresponding hours of $A$. To predict a test matrix value $B(i,j)$, the load of consumer $j$ at hour $i$, we identify hours of $A$ (represented by clusters in ${\cal C}$) that are {\em `most similar} to the query hour $i$ and report an average of loads of consumer $j$ and that of its top $k$ neighbors in those clusters. 
The only information of a query hour we can use is its calendar attributes, i.e., time, day, and month. On the other hand, attributes of clusters of training hours (columns of $H$) are based on overall consumption at those hours. 


For a common representation of an hour and a cluster of hours, we use a $75$-dimensional vector, $\vect{v}(\cdot)$. The first $24$ coordinates of $\vect{v}(\cdot)$ ($\vect{v}[0\ldots 23]$) correspond to the $24$ hours in a day. The next $7$ coordinates represent the days of a week, and the $31$ coordinates after them represent the days of a month. The following $12$ coordinates stand for the months of a year. The $74^{th}$ and $75^{th}$ coordinates encode public holidays. 

\begin{figure}[ht!]
	\centering
	\includegraphics[page=19, width=\linewidth]{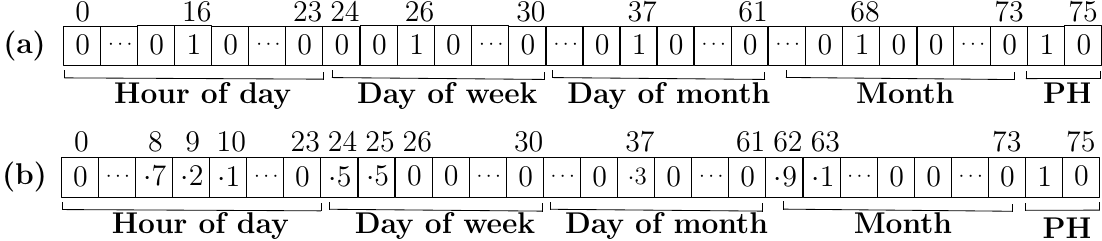}
	\caption{Vector encoding of (a) an hour: Mon $7^{th}$ Jul, 4-5 PM (b) a cluster containing $\mathbf{70\%}$ 8 AM, $\mathbf{20\%}$ 9 AM and $\mathbf{10\%}$ 10 AM hours, half each of  Mondays and Tuesdays. $\mathbf{90\%}$ hours are of Jan and $\mathbf{10\%}$ of Feb.}
	\label{vectorRep}
\end{figure}

For a given hour $h$ (a time-stamp with associated calendar information), the value of $\vect{v}(h)$ at a coordinate is $1$ if $h$ has the corresponding attribute. Figure~\ref{vectorRep}(a) depicts an example vector representation of an hour. For a cluster $C = \{h_1,h_2,\ldots,h_{|C|}\}$ of hours, $\vect{v}(C)$ is the distribution of hours (represented as $\vect{v}(\cdot)$) contained in $C$. Formally, $\vect{v}(C) = \frac{1}{\vert C \vert} \sum_{h \in C} \vect{v}(h).$ Figure~\ref{vectorRep}(b) depicts an example vector representation of a cluster of hours (rows of $H$).

\subsection{Forecasting the Load} 
Finally, we describe the mechanism to estimate load of household $j$ at a testing hour $i$, i.e. value of $B(i,j)$. To predict $B(i,j)$, we report an `average' of household $j^{th}$'s load along with the load of its top $k$ neighbors in the clusters of hours in $A$ that are most similar to the query hour $i$. For this, we need a measure of distance/similarity, $d({h},C)$ between a testing hour ${h}$ and a cluster of training hours $C$. Since, the vector representations $\vect{v}({h})$ and $\vect{v}(C)$ (for hour $h$ and cluster$C$ respectively), encode attributes of $h$ and $C$ related to time, day and month etc. We define $d({h},C)$ to be a weighted sum of the $l_{p}$-distance between the corresponding attributes of ${h}$ and $C$.
More precisely, let $d[q]$ be the absolute difference between the $q^{th}$ coordinate of $\vect{v}({h})$ and $\vect{v}(C)$, i.e. $d[q] = \vert \vect{v}({h})[q] - \vect{v}(C)[q] \vert$. Consider a vector $``\vect{w}"$ that contains weights of attributes $w_{1}, w_{2}, \ldots, w_{8}$.
Also consider another vector $``\vect{a}"$ that contains following elements of the feature vector related to calendar attributes: $a_{1} = \bigg( \sum\limits_{q=0}^{23} (d[q])^{p} \bigg)^{\nicefrac{1}{p}}$, $a_{2} = \bigg(\sum\limits_{q=24}^{30} (d[q])^{p} \bigg)^{\nicefrac{1}{p}}$, $a_{3} = \bigg(\sum\limits_{q=31}^{61} (d[q])^{p} \bigg)^{\nicefrac{1}{p}}$, $a_{4}=\bigg(\sum\limits_{q=62}^{73} (d[q])^{p} \bigg)^{\nicefrac{1}{p}}$,\\ $a_{5}=\bigg(\sum\limits_{q=74}^{75} (d[q])^{p} \bigg)^{\nicefrac{1}{p}}$. 

The distance $d(h,C)$ between $h$ and $C$ is defined as $d(h,C) := \vect{a^{T}} \vect{w}$.
The optimum values for the weights of attributes ($w_1, w_2, \ldots , w_8$) are computed using the standard validation set approach~\cite{validationSetApproach}.
The similarity, $sim(h,C)$ between an hour ${h}$ and a cluster $C$ is given by
$	sim({h},C) := 1-d({h},C)$.
To predict $B(i,j)$, we find a subset ${\cal C}' \subset {\cal C}$ of $t$ clusters that have the highest similarity with the hour $i$ and a subset of top $k$ nearest neighbors of consumer $j$ (i.e. $N_{1}(j), \cdots,  N_{k}(j)$). We report the similarity-weighted mean of the median consumption of sub-matrix, which consists of loads of consumer $j$, $N_{1}(j), \cdots , N_{k}(j)$ in these $t$ clusters (we empirically select $t=2$ for all datasets). 
In other words, \textsc{fmf} computes a forecast $B'(i,j)$ for the load $B(i,j)$ as follows:

\begin{equation}
  B'(i,j) =  \frac{\sum_{C \in \mathcal{C'}} \{sim(i,C) \times \textsc{median}_{h \in C} \cdot \{a \} \}}{\sum_{C \in \mathcal{C'}} sim(i,C)}
\end{equation}
\begin{equation}
    a \;=\; \textsc{median}_{x \in N(j)} A(h,x)
\end{equation}

\subsection{Time Complexity of \textsc{fmf}} \label{runtimeAnalysisSection}
The running time of the preprocessing is linear with respect to the size of the input. The training matrix $A$ has dimensions $m_{1} \times n$, \textsc{svd} on $A$ that takes $O(\min(m_{1} n^{2}, m_{1}^{2} n))$ time. The next step of the algorithm is clustering training hours. The standard $k$-means algorithm takes time proportional to $O(n r I)$, where $r$ is the number of clusters and $I$ is the number of iterations of the $k$-means algorithm (similar is the case for clustering the customers). Note that these $t$-nearest neighbors are found for each hour only, not for individual consumers, i.e., for every row of $B$, we perform this step only once. Therefore, the total time required for all nearest neighbors computations is $O(m_2rt)$. Thus the total running time of all these steps is $O(\min(m_{1} n^{2}, m_{1}^{2} n) + n r I + m_2rt )$, where $r,t,I$ are user-set parameters and usually small constants. Hence the time complexity is dominated by the \textsc{svd} step. To make a forecast, we compute the medians in each nearest cluster and report an average of these medians. Thus the worst-case runtime of a forecast is $O(m_1)$.

\section{Experimental Setup}\label{Experiments}
In this section, we first describe the three benchmark datasets that we use for evaluating \textsc{fmf}. We then describe the evaluation metrics used to measure the goodness of our approach. We also discuss the baseline and state-of-the-art methods for \textsc{stlf} used for comparison with \textsc{fmf}. In the end, we show the visual representation of data to analyze the hidden patterns (if any exist in the data). Our algorithms are implemented in Matlab and Python on a Core i7 \textsc{pc} with 8GB memory. Code of \textsc{fmf} and the pre-processed datasets are available online\footnote{\url{https://github.com/sarwanpasha/Load-Forecasting}} for reproducibility.

\subsection{Dataset Description and Visualization} \label{Dataset_detail}
We use real-world smart meter data of hourly consumption from different residential areas of Sweden~\cite{javed2012forecasting,ali2019short,sarwan2019soft},  Australia~\cite{lusis2017short} and Ireland~\cite{irish_dataset}. Table~\ref{tbl:Dataset} shows detail statistics of these datasets. 

\begin{table}[ht!]
	\centering

	\begin{tabular}{p{0.8cm}cp{0.6cm}p{2.7cm}p{0.4cm}p{0.4cm}} 
		\toprule
		{Dataset} & {Households} & {Hours} & {Duration} & Avg. load & Std. dev.\\ [0.5ex] 
		\midrule
		{Sweden} & {582} & {17544} & Jan 1,\ 04-Dec 31,\ 05 & {2.52} & {0.81} \\ 
		\midrule
		{Ireland} & {709} & {12864} & Jul 14,\ 09-Dec 31,\ 10 & {1.33} & {1.33}  \\ 
		\midrule
		{Australia} & {34} & {26304} &  Jul 1,\ 10-Jun 30,\ 13 & {0.79} & {0.29}  \\ 
		\bottomrule
	\end{tabular}
		\caption{Statistics of datasets: Only the first $12864$ hours ($\approx 18$ months) data is used from all datasets.}
	\label{tbl:Dataset}
\end{table} 

To visually examine natural clusters in the data (if any), we embed time-stamps into a $2d$ real vector space using $t$-distributed stochastic neighbor embedding ($t$-\textsc{sne})~\cite{van2008visualizing}. Recall that $X \in \mathbb{R}^{m \times n}$ is the load matrix,  where $m$ is the number of hours, and $n$ is the number of households. We apply $t$-\textsc{sne} on the rows of $X$ to get a matrix $F \in \mathbb{R}^{m \times 2}$. We plot the rows of $F$ with each row labeled based on the calendar attributes to observe patterns in the data visually.

\begin{figure}[h!]
\centering
\begin{subfigure}{.25\textwidth}
  \centering
  \includegraphics[width=\linewidth] {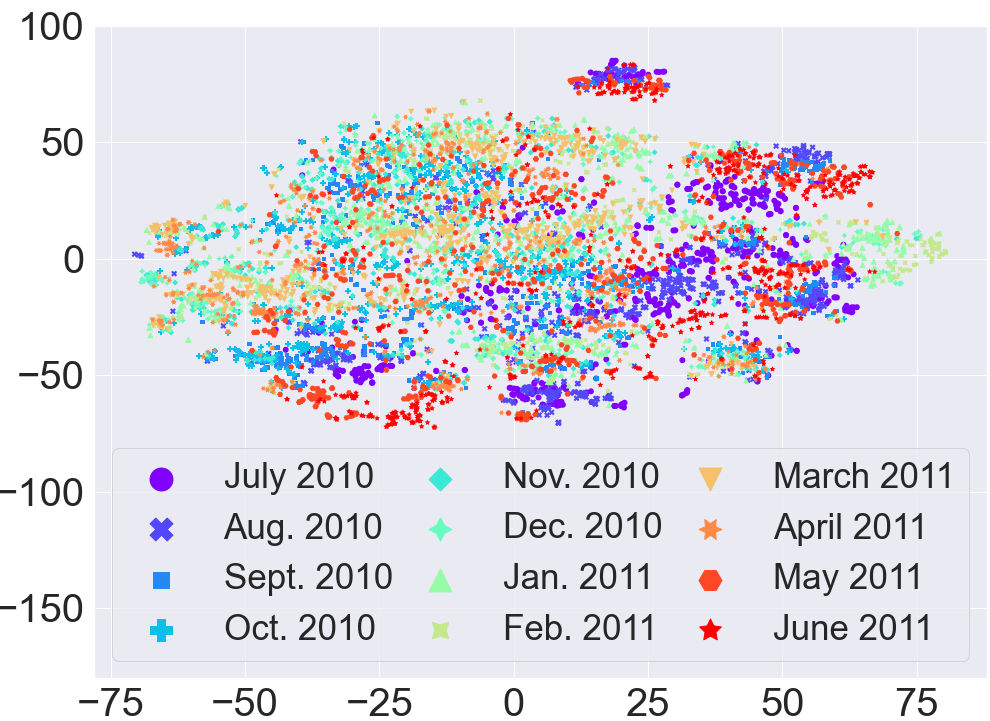}
  \caption{Month wise labels}
  \label{fig_tsne_australia_month_wise}
\end{subfigure}%
\begin{subfigure}{.25\textwidth}
  \centering
  \includegraphics[width=\linewidth] {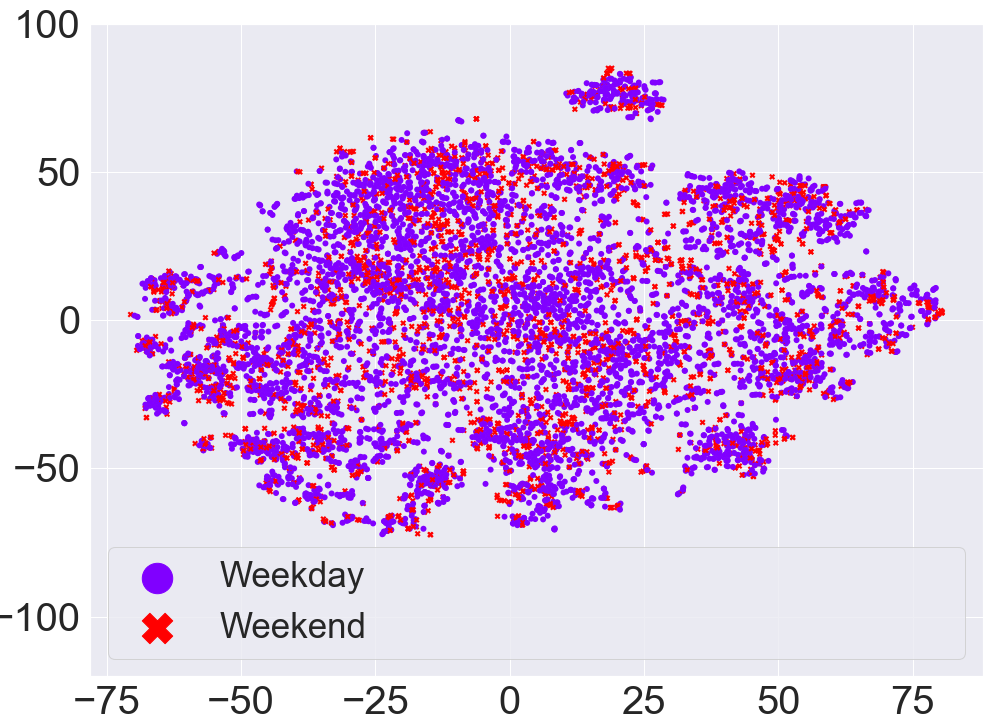}
  \caption{Weekdays/Weekends}
  \label{fig_tsne_australia_weekdays_weekends}
\end{subfigure}%
\\
\begin{subfigure}{.25\textwidth}
  \centering
  \includegraphics[width=\linewidth] {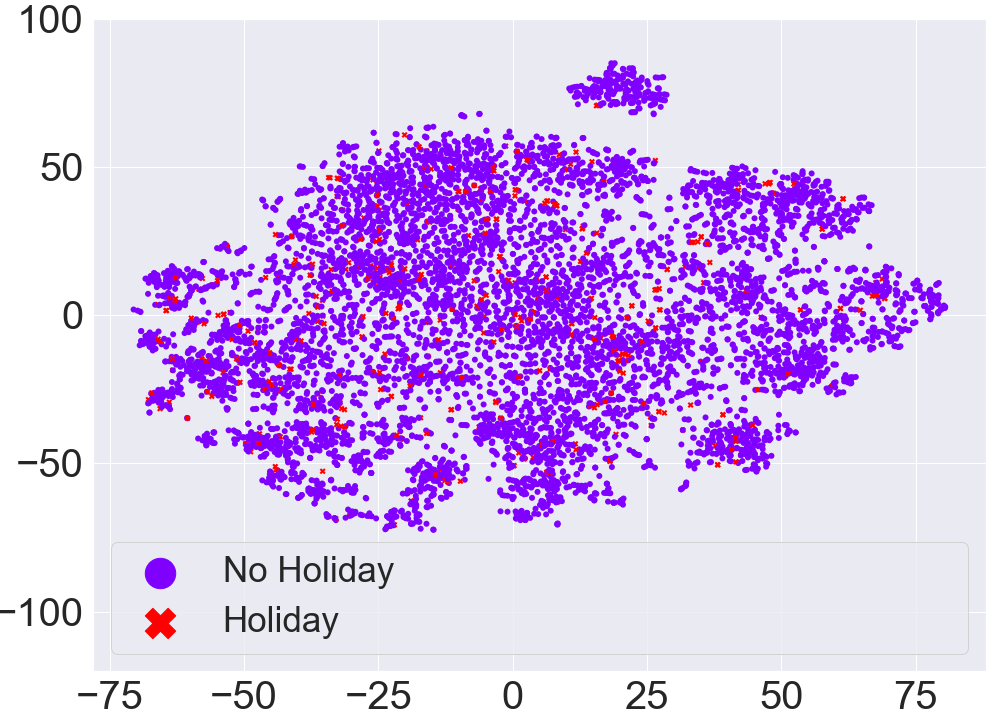}
  \caption{Public Holiday}
  \label{fig_tsne_australia_public_holiday}
\end{subfigure}%
\begin{subfigure}{.25\textwidth}
  \centering
  \includegraphics[width=\linewidth] {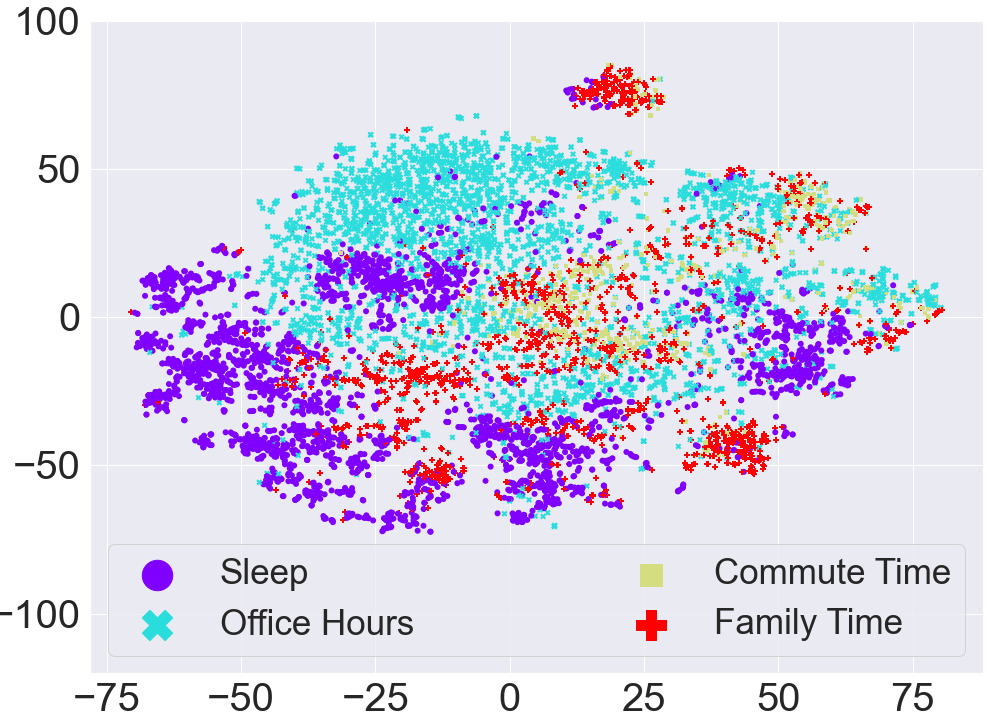}
  \caption{Hours of the Day}
  \label{fig_tsne_australia_hours}
\end{subfigure}
\caption{The t-\textsc{sne} plots of different labels for Australia dataset. Figure is best seen in color.}
\label{fig_tsne_australia_labels}
\end{figure}

Figure~\ref{fig_tsne_australia_month_wise} shows the $t$-\textsc{sne} plot for the Australia dataset with hours assigned month names as labels. Although we can observe some small clusters (i.e., July 2010 and June 2011 at the top center of the plot), there is no clear separation between data points based on months. Similarly, there is no clear separation between hours based on ``Weekdays'' and ``Weekends'' or based on  ``Public Holiday'' and ``No Public Holiday'' (Figure~\ref{fig_tsne_australia_weekdays_weekends} and~\ref{fig_tsne_australia_public_holiday}). We also group hours of a day into $4$ periods namely ``Sleep Time" (12:00 AM -- 8:00 AM), ``Office Hours" (8:00 AM -- 5:00 PM), ``Commute Time" (5:00 PM -- 7:00 PM), and ``Family Time" (7:00 PM -- 12:00 AM). We can observe in Figure~\ref{fig_tsne_australia_hours} that there are some patterns for sleep hours, office hours, and family time. However, there is no complete separation between different labels. Overall, this scattered behavior shows no clear pattern in the data, i.e.,  extracting any useful information from data without any preprocessing is not easy. 




The $t$-\textsc{sne} plots for the Sweden dataset, Figure~\ref{fig_tsne_swe_labels} (in appendix), unlike the Australia dataset, reveal some grouping between months ((Figure~\ref{fig_tsne_swe_month_wise}) and days (weekend/weekdays Figure~\ref{fig_tsne_swe_weekdays_weekends}). In the Ireland dataset, we can observe that the weekends and weekdays are clearly grouped (Figure~\ref{fig_tsne_ire_weekdays_weekends}). Similarly, the hours of the day in Figure~\ref{fig_tsne_ire_hours}  show a separation between office hours and sleep hours, but the family time and commute time overlap.

\subsection{Evaluation Setup and Metrics}
We use the following error metrics to test the performance of \textsc{fmf}: Mean Absolute Error (\textsc{mae})~\cite{shi2018deep}, Mean Absolute Percentage Error (\textsc{mape})~\cite{kong2019short}, Root Mean Square Error (\textsc{rmse})~\cite{lusis2017short} and Normalized Root Mean Square Error (\textsc{nrmse}).
From all datasets, we use data of first $18$ months in our experiments. Out of these $18$ months data, the first $\approx 12$ months of data ($m_1 = 8760$ (hours))  is used as training data. The succeeding $\approx 6$ months data ($m_{2} = 4104$ (hours)) is used for testing.

\subsubsection{Clustering Evaluation}
We evaluate the effectiveness of clustering hours in the abstract feature-space (rows of $H = U_d\Sigma_d$) by observing clusters representations, $\vect{v}(\cdot)$. We note that hours expected to have similar consumption based on domain knowledge tend to be grouped into the same clusters. 
\begin{figure}[ht!]
	\centering
	\includegraphics[page=19, width=\linewidth] {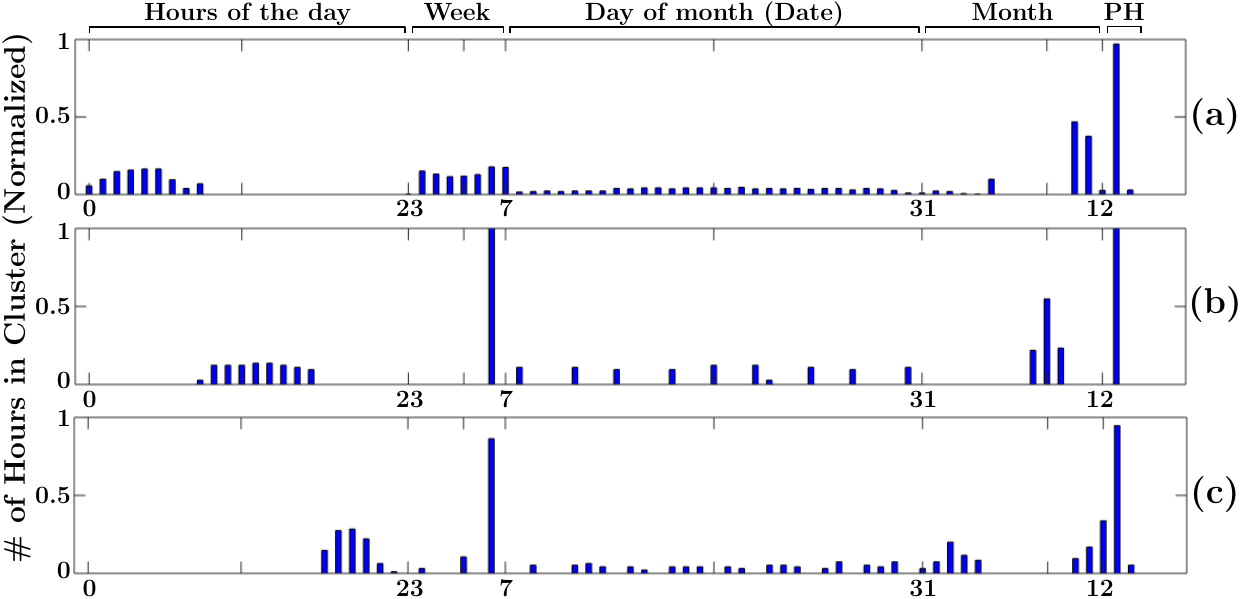}
	\caption{Bar graphs of $\vect{v}(\cdot)$ for three clusters of hours (Ireland dataset). All three clusters (a), (b), and (c) shows that similar load pattern hours tend to be in the same clusters.}
	\label{cluster_efficiency_irish}
\end{figure}
Figure~\ref{cluster_efficiency_irish}  depicts three randomly chosen clusters and shows that clustering of rows of $H$ is meaningful and successfully avoids the curse of dimensionality. The first cluster (a) contains the winter night hours ($12 AM$ to $8 AM$) of one whole week. The second cluster (b) is for the daytime summer weekends, while the last cluster (c) is for the evening hours of winter weekends. 
\begin{remark}
Note that we perform clustering only once, and the same clusters are used to predict the entire test matrix $B$ (i.e. $\approx 6$ months hourly load). 
\end{remark}

Figure~\ref{fig_clust_goodness_swe} (in appendix) shows the clusters of consumers for Sweden dataset. Clusters are well separated in terms of load, highlighting the effectiveness of clustering in our representation. Similar behavior is observed for other datasets.

\subsection{Comparison Algorithms}
We compare the results of \textsc{fmf} with several baselines and state-of-the-art methods proposed in the literature.

\subsubsection{Baseline Method}
We use Auto Regressive Integrated Moving Average (\textsc{arima}), a time series model, as a baseline~\cite{shi2018deep}. \textsc{arima} treats historical loads as a time series and attempts to learn parameters for forecasting future values. Our second baseline is {\em `Persistent Forecast'}, which uses an average of ``previous hours'' loads as the forecast for the next hour.

\subsubsection{State-of-the-Art (\textsc{sota}) Methods}
The \textsc{sota}  methods that we use for comparison with \textsc{fmf} are the following: Long Short Term Memory (\textsc{lstm})~\cite{kong2019short,kell2018segmenting}, Multiple Linear Regression (\textsc{mlr})~\cite{lusis2017short}, Regression Trees (\textsc{rt})~\cite{lusis2017short}, 
Neural Network (\textsc{nn})~\cite{lusis2017short,kell2018segmenting}, Support Vector Regression (\textsc{svr})~\cite{kell2018segmenting}, and Random Forest (\textsc{rf})~\cite{kell2018segmenting}. Further implementation details (including hyperparameters values) are given in the appendix.




We also compare \textsc{fmf} with the methods proposed in~\cite{lusis2017short} and ~\cite{kell2018segmenting}. In~\cite{lusis2017short}, the authors reported household level \textsc{stlf} results on the Australia dataset using four different machine learning algorithms. We apply \textsc{fmf} on the same dataset (with the same train-test split and settings) and report forecasting accuracy (using the same error metrics). In~\cite{kell2018segmenting}, the authors employed four machine learning models to forecast loads for clusters of consumers for the Ireland dataset. We use their method for clustering consumers into clusters and forecast their loads using \textsc{fmf} and compare the forecasting error with the best and recommended method by~\cite{kell2018segmenting}.

\section{Results and Discussion} \label{results_and_comparisons}
In this section, we report forecasting results of \textsc{fmf} perform a comparison with the baseline, and \textsc{sota}  approaches. Since \textsc{fmf} works at any level of user-defined granularity both in spatial and time domains, we demonstrate the effectiveness of \textsc{fmf} for predicting aggregated load of clusters of households and total load for the extended periods ranging from a couple of hours to days.

\subsection{Evaluation of  Hourly forecast at Household level}
\begin{figure}[ht!]
	\centering
	\includegraphics[page = 16,width=\linewidth] {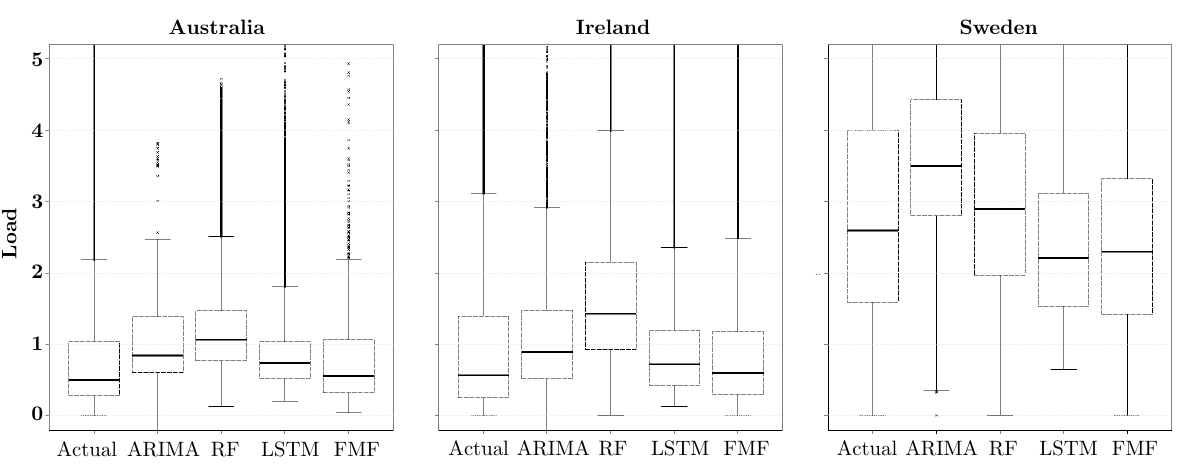}
	\caption{Actual and forecasted loads of all hours and all households  for \textsc{arima}, \textsc{rf}, \textsc{lstm} and  \textsc{fmf}. Observe that \textsc{fmf} forecasts are significantly closer to actual loads in all datasets.}
	\label{actual_vs_pred_load_comparison_all_aus}
\end{figure}

In this section, we use \textsc{fmf} to perform \textsc{stlf} at household level, i.e. we forecast individual entries of the test matrix $B$. Figure~\ref{actual_vs_pred_load_comparison_all_aus} shows the boxplots of actual hourly loads (all $m_2\times n$ values in $B$) and the hourly loads forecasted using \textsc{arima}, \textsc{rf}, \textsc{lstm}, and \textsc{fmf}.  Note that the loads predicted using \textsc{fmf} is closer to the actual loads than all other methods in all datasets.

\noindent Table~\ref{result_table_row_wise} shows the comparison of different methods for hourly load prediction with \textsc{fmf} in terms of average error over all households and hours. In this experimental setting, we forecast the hourly load for the next $6$ months. Note that the actual load values in all datasets are very close to $0$ (see Table~\ref{tbl:Dataset}), since they represent individual household loads for a short duration (an hour). Thus, a slight prediction error results in a higher percentage error. This is demonstrated by higher \textsc{mape} ($>30\%$) for all methods. However, \textsc{rmse}, \textsc{nrmse}, and \textsc{mae} values in Table~\ref{result_table_row_wise} are small, indicating that the predicted loads are close to actual loads (which is also evident from Figure~\ref{actual_vs_pred_load_comparison_all_aus}). 
\begin{table}[ht!]
	\centering
	\begin{tabular}{p{1.8cm}p{1.1cm}p{0.8cm}p{0.9cm}p{1.2cm}p{0.65cm}}
		\toprule
		\multirow{2}{*}{Method} & \multirow{2}{*}{Dataset}  & \textsc{mae} {\em (kWh)} & \textsc{rmse} {\em (kWh)} & \textsc{nrmse} & \textsc{mape} {\em (\%)} \\ [0.5ex] 
		\midrule
		\multirow{ 3}{*}{\textsc{arima}}  & Sweden & 1.55 & 1.87 & 10.79  & 98.5\\
		& Ireland & 1.01 & 1.44 & 48.50 & 223.2\\
		& Australia & 0.71 & 0.89 & 6.79 & 199.5 \\ 
		\midrule
		\multirow{ 3}{*}{\textsc{rf}~\cite{kell2018segmenting}} & Sweden & 1.03 & 1.57 & 0.018 & 51.49  \\
		& Ireland & 1.20 & 2.06 & 0.02 & 354.9 \\
		& Australia & 0.72 & 0.98 & 0.09 & 234.2 \\ 
		\midrule
		\multirow{ 3}{*}{\textsc{lstm}~\cite{kong2019short}}   & Sweden & 0.79 & 1.35 & 0.017 & \textbf{31.85}  \\
		& Ireland & 0.66 & 1.30 & 0.015 & 149.9 \\
		& Australia & 0.405 &  0.65 & 0.06 & \textbf{90.68}\\ 
		\midrule
		\multirow{ 3}{*}{\textsc{fmf}} & Sweden & \textbf{0.78} & \textbf{1.24} & \textbf{0.012} & 37.1  \\
		& Ireland &  \textbf{0.60} & \textbf{1.20} & \textbf{0.014} & \textbf{92.9} \\
		& Australia & \textbf{0.401} &  \textbf{0.60} & \textbf{0.05} & 97.4 \\ 
		\midrule
		\multirow{3}{2cm}{\textsc{fmf} $(\%)$ improvement over \textsc{arima}} & \multirow{1}{*}{Sweden}  & \multirow{1}{*}{$49.67$}  & \multirow{1}{*}{$33.68$}  & \multirow{1}{*}{$99.9$}  & \multirow{1}{*}{$62.3$}\\
		\cline{2-6}
		 & \multirow{1}{*}{Ireland}  & \multirow{1}{*}{$40.59$}  & \multirow{1}{*}{$16.66$}  & \multirow{1}{*}{$99.9$}  & \multirow{1}{*}{$58.3$}\\
		 \cline{2-6}
		 & \multirow{1}{*}{Australia}  & \multirow{1}{*}{$43.6$}  & \multirow{1}{*}{$32.58$}  & \multirow{1}{*}{$99.2$}  & \multirow{1}{*}{$51.1$}\\
		\bottomrule
	\end{tabular}
		\caption{Per household average errors of all hours for the next $6$ months load forecasting.}
\label{result_table_row_wise}
\end{table}
\begin{remark}
Note that we are forecasting hourly load for the next six months. This is still called short term load forecasting as the hourly load is being forecasted. In the case of medium term load forecasting, the ``total load" for a few months is forecasted rather than the hourly load.
\end{remark}

\begin{figure*}[hbt!]
	\centering
	\begin{subfigure}{.32\textwidth}
		\centering
		\includegraphics[ width=\linewidth]{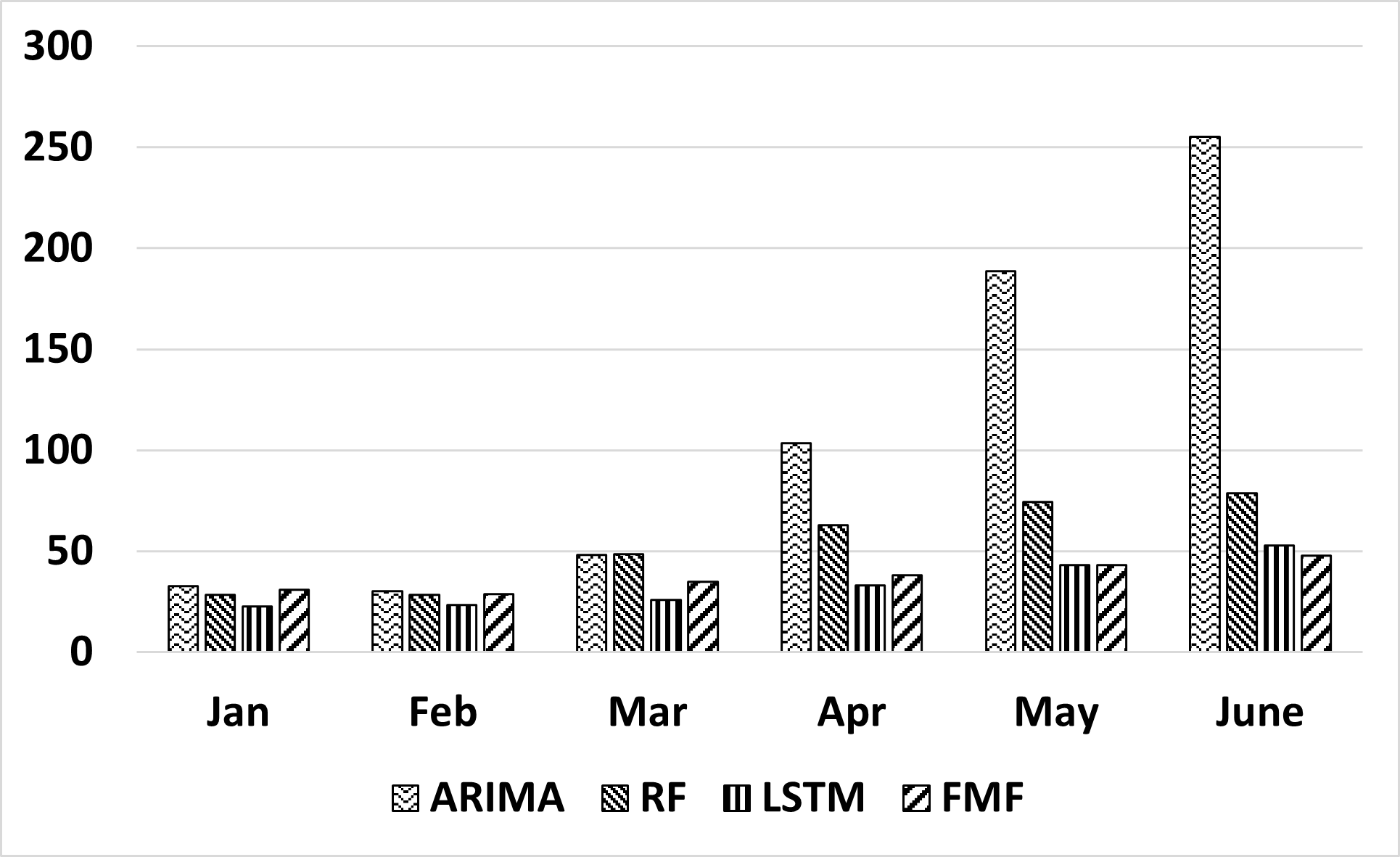}
    	\caption{}
        \label{fig_month_wise_mape_swe}
	\end{subfigure}%
	\hspace{0.05cm}
	\begin{subfigure}{.32\textwidth}
		\centering
		\includegraphics[ width=\linewidth]{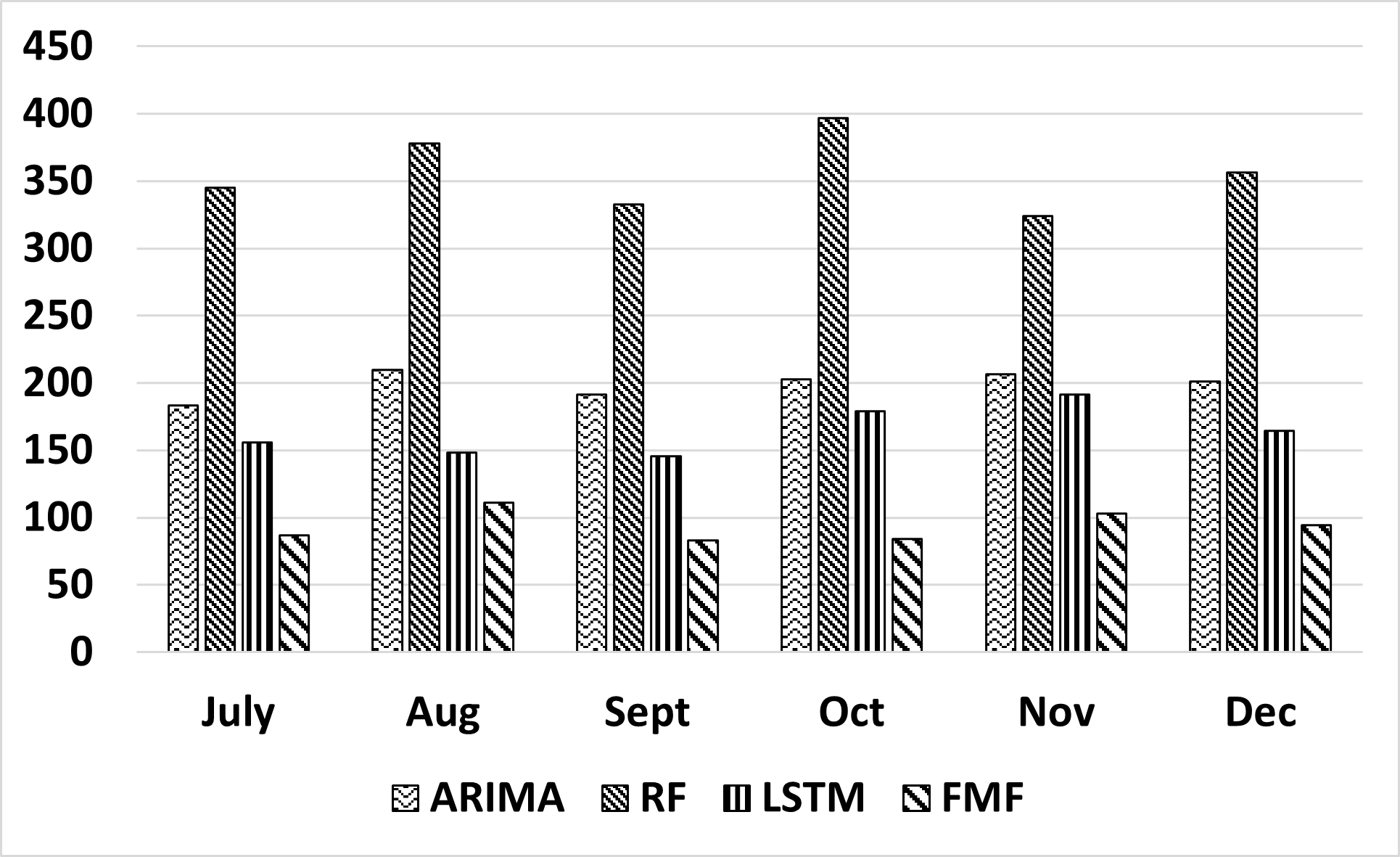}
		\caption{}
        \label{fig_month_wise_mape_ire}
	\end{subfigure}
		\hspace{0.03cm}
	\begin{subfigure}{.32\textwidth}
		\centering
		\includegraphics[ width=\linewidth]{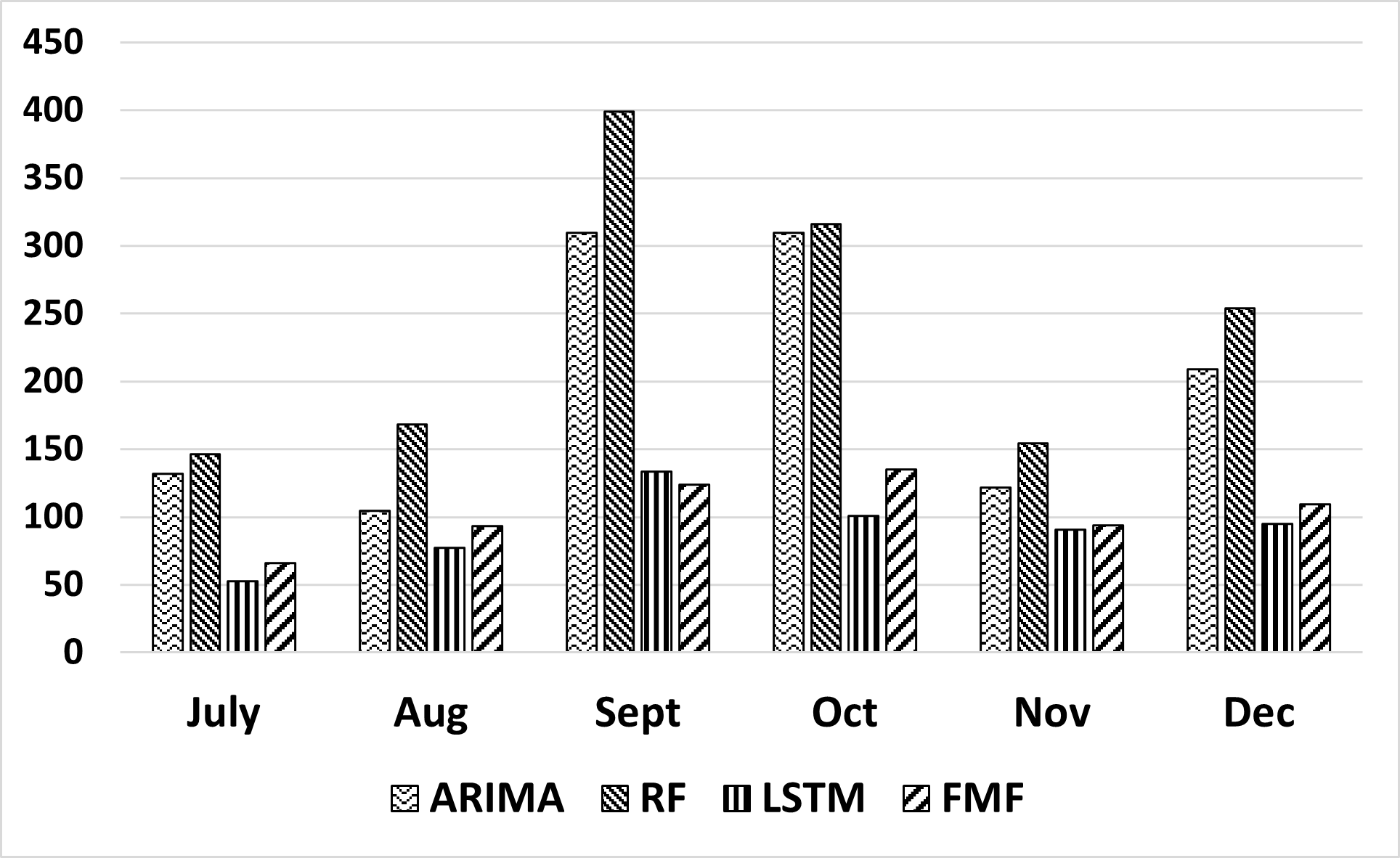}
        \caption{}
        \label{fig_month_wise_mape_aus}
	\end{subfigure}
	\caption{Monthly \textsc{mape} of \textsc{arima}, \textsc{rf}, \textsc{lstm} and  \textsc{fmf} for (a) Sweden (b) Ireland (c) Australia  dataset (lower value is better). }
    \label{fig_month_wise_mape_all_data}
\end{figure*}




Since variability in low loads is relatively much higher, algorithms tend to perform better if the load values of households are on the higher side. Table~\ref{tbl:Dataset} shows the average and standard deviation of loads of three datasets. The average load is the largest for the Sweden dataset, and we achieve the minimum \textsc{mape} on it compared to other datasets (see Table~\ref{result_table_row_wise}). The average load of households for the Australia dataset is the smallest among the three datasets, and we got a larger \textsc{mape} for this dataset. This also points to the widely accepted phenomenon that forecasting larger aggregated loads is substantially easier than smaller individual loads. We observed in Figure~\ref{fig_tsne_australia_labels}, Figure~\ref{fig_tsne_swe_labels} (in appendix), and Figure~\ref{fig_tsne_ire_labels} (in appendix) that the Sweden and Ireland datasets are better separated by different calendar attributes, compared to the Australia dataset. This separability leads to better results on the former two datasets by all error measures.

\begin{figure}[ht!]
	\centering
	\includegraphics[page = 17,width=\linewidth] {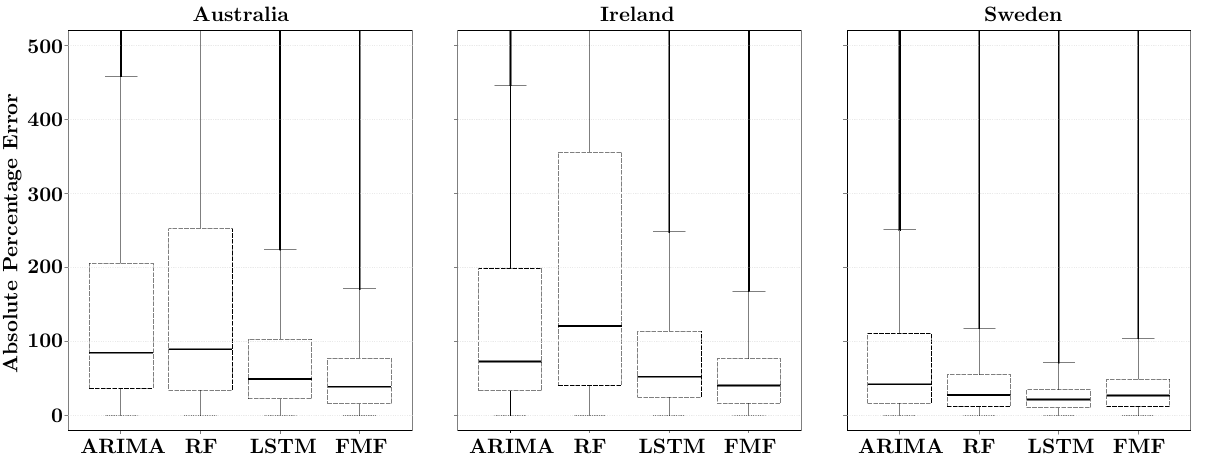}
	\caption{Absolute percentage errors of all households and all hours for \textsc{arima}, \textsc{rf}, \textsc{lstm}, and  \textsc{fmf}.}
	
	\label{actual_vs_pred_load_MAPE_comparison_all_aus}
\end{figure}

From Table~\ref{result_table_row_wise} it is clear that in most cases, \textsc{fmf} outperforms the baseline and \textsc{sota}  approaches. However, these are the mean values of errors, which are very sensitive to outliers. Recall that many actual loads are $0$ or very close to it; there is a sizeable number of outliers in the errors. Therefore, we also plot all point-wise absolute percentage errors in Figure~\ref{actual_vs_pred_load_MAPE_comparison_all_aus} to show that for majority of consumers \textsc{fmf} achieves very low percentage errors. It is clear from Figure~\ref{actual_vs_pred_load_MAPE_comparison_all_aus} that the median percentage error achieved by \textsc{fmf} is the smallest compared to other methods (on the Sweden dataset it is very close to the median performance of \textsc{lstm}).

\subsubsection{Month-Wise Performance Comparison}
To analyze the effect of seasons on forecasting, we report the results for different months. We observe that on the Sweden dataset (Figure~\ref{fig_month_wise_mape_swe}), \textsc{fmf} is better than all methods for May and June, better than \textsc{arima} and \textsc{rf} for all other months, and comparable to \textsc{lstm} for Jan to  April. For the Ireland and Australia dataset (Figure~\ref{fig_month_wise_mape_ire} and Figure~\ref{fig_month_wise_mape_aus}), \textsc{fmf} is better than all methods for every month, except in the Australia dataset, it is comparable with \textsc{lstm}.



\subsubsection{Comparison with Method Proposed in~\cite{lusis2017short}}
Authors in~\cite{lusis2017short} perform \textsc{stlf} at even finer granularity ($30$ minutes ahead) using different machine learning methods to perform \textsc{stlf}. For comparison, as performed in~\cite{lusis2017short}, we use $28$ days for testing and all remaining data for training ``$\approx 2.5$ years" (for Australia dataset). The test data consists of randomly selected weeks in September ($2012$), December ($2012$), March ($2013$), June ($2013$). 
The comparison, using same parameters, dataset and experimental settings as in~\cite{lusis2017short}, is given in Table~\ref{tbl_comparisonLusis}. Observe that \textsc{fmf} achieves up to $26.5\%$ improvement in  \textsc{rmse} and $94.1\%$ improvement in \textsc{nrmse}.

\begin{table}[ht!]
	\centering
	\begin{tabular}{lcc}
		\toprule
		Method & \textsc{rmse} {\em (kWh)} & \textsc{nrmse} \\
		\midrule
		\textsc{mlr}  & 0.561 & 2.778  \\
		\textsc{rt} & 0.516 &  2.375  \\
		\textsc{svm}  & 0.531 & 1.793 \\ 
		\textsc{nn}  & 0.53 & 2.772\\
		\textsc{fmf} & \textbf{0.390} & \textbf{0.140} \\
		\midrule
		\textsc{fmf} (\%) improvement over \textsc{rt} & \multirow{1}{*}{$24.4$}  & \multirow{1}{*}{$94.1$} \\
		\textsc{fmf} (\%) improvement over \textsc{svm}  & \multirow{1}{*}{$26.5$}  & \multirow{1}{*}{$92.1$} \\
		\bottomrule
	\end{tabular}
		\caption{Comparison of \textsc{fmf} with different machine learning methods employed by~\cite{lusis2017short} for Australia dataset.}
	\label{tbl_comparisonLusis}
\end{table}

\subsubsection{Comparison with Persistent Forecast and Autocorrelation Analysis}
We designed two settings for the persistent forecast (\textsc{pf}) model and compared them with the \textsc{fmf} results. In the first setting, \textsc{pf}$_{1}$, given a query hour $t$, we take the load values for $t-1$, $t-2$ hours of the same day, and $t$ and $t-1$ hours of the previous day and take the average of all these load values as the predicted value. In the second setting, \textsc{pf}$_{2}$, given a query hour $t$, we take the load values for $t-1$, $t-2$ hours of the same day, $t$ and $t-1$ hours of the previous day, and $t$ and $t-1$ hours of the same day of the previous week and take the average of all these load values as the predicted value. The result comparison for the \textsc{pf} with \textsc{fmf} are shown in Table~\ref{tbl_pf_aus}. A limitation of \textsc{pf} is that it can only be used to forecast the load for the next hour and cannot forecast the hourly load for multiple hours in advance (e.g., hourly load for the next six months or long-term load (e.g., the aggregated load of a month or year).  

\begin{table}[ht!]
	\centering
	\begin{tabular}{ccccc}
		\toprule
		Dataset & Method & \textsc{rmse} {\em (kWh)} & \textsc{nrmse} & \textsc{mape} (\%) \\
		\midrule
		 \multirow{ 3}{*}{Australia} & \textsc{pf}$_{1}$ & 0.34 & 0.15 &  \textbf{55.53} \\
		 & \textsc{pf}$_{2}$ & 0.46 & 0.20 & 65.83 \\
		 & \textsc{fmf} & \textbf{0.33} & \textbf{0.14} & 117.81 \\ 
		\midrule
		 \multirow{ 3}{*}{Sweden} & \textsc{pf}$_{1}$ & 0.99 & 0.037 & 30.17   \\
		 & \textsc{pf}$_{2}$ & 1.01 & 0.038 & 31.57 \\
		 & \textsc{fmf} & \textbf{0.95} & \textbf{0.035} & \textbf{26.15} \\ 
		\midrule
		 \multirow{ 3}{*}{Ireland} & \textsc{pf}$_{1}$ & 0.68 & 0.041  & 102.61  \\
		 & \textsc{pf}$_{2}$ & 0.59 & 0.036 & 99.42 \\
		 & \textsc{fmf} & \textbf{0.58} & \textbf{0.034} & \textbf{64.23} \\
		\bottomrule
	\end{tabular}
		\caption{Forecasting errors of \textsc{pf}$_{1}$, \textsc{pf}$_{2}$, and \textsc{fmf}.}
	\label{tbl_pf_aus}
\end{table}



We use autocorrelation to measure the linear relationship between lagged values. We consider lag values (time gap) ranging from $1$ to $20$. A lag $k$ autocorrelation is the correlation between values that are $k$ periods apart. There is a very low average autocorrelation with a small standard deviation in the Sweden dataset (Figure~\ref{fig_corr_swe} in appendix). These values are relative larger in the Australia dataset (Figure~\ref{fig_corr_aus} in appendix). The Ireland dataset has autocorrelation on the lower end too. This explains why \textsc{pf} results on the Australia data are better than the Sweden dataset, hence using more lag values would not increase the forecasting accuracy of \textsc{pf}. 





\subsubsection{Effectiveness of \textsc{svd}}
We showed the effectiveness of \textsc{svd} in preserving the original structure of data and making clear clusters of hours (see Section~\ref{Dataset_detail}). Now we report \textsc{mape} for \textsc{fmf} with and without using \textsc{svd}. We observe that \textsc{svd} helps improve the forecasting error for all datasets.

\begin{table}[ht!]
	\centering
	\begin{tabular}{lccc}
		\toprule
		Method & Australia & Sweden & Ireland \\
		\midrule
		\textsc{fmf} without \textsc{svd} & 104.79 & 38.85  & 103.81  \\
		\textsc{fmf} with \textsc{svd} & \textbf{97.4} & \textbf{37.1}  & \textbf{92.9}  \\
		\bottomrule
	\end{tabular}
		\caption{\textsc{mape} (for the hourly forecast of 6 months) for \textsc{fmf} with and without using \textsc{svd} for hours representation.}
	\label{tbl_svd_comparison}
\end{table}

Moreover, \textsc{svd} also helps reduce the computational runtime of the underlying clustering algorithm (by reducing the dimensionality of the data). We report the effect of dimensionality reduction on the running time of clustering $m$ hours. Recall that each hour is an $n$-d vector ($n=709$, the number of households in the Ireland dataset). Figure~\ref{fig_clust_dim_time} (in appendix) plots the runtimes clustering varying number of hours through $k$-means algorithm ($k=80$), when dimensionality of hours is $709$ and $424$ (approximately $80\%$ energy in the singular values is preserved, see Section~\ref{section_Matrix_Factorization}).   



\subsection{Evaluation of \textsc{fmf} at Higher Granularity}

As discussed above, load forecasting at coarser levels (either in space or time dimensions) is substantially easier. This, however, is an important problem in the power sector decision-making. \textsc{fmf} is most suited for \textsc{stlf} at the individual household level, but it is a general-purpose method that works at any level of user-defined granularity and yields significantly more accurate forecasts in most cases. In this section, we use \textsc{fmf} to forecast load for longer durations (up to a day), for groups (clusters) of households, and both combined.
    
\subsubsection{Forecast for longer durations}
\begin{figure}[h!]
\centering
\begin{subfigure}{.33\linewidth}
  \centering
  \includegraphics[width=\linewidth]{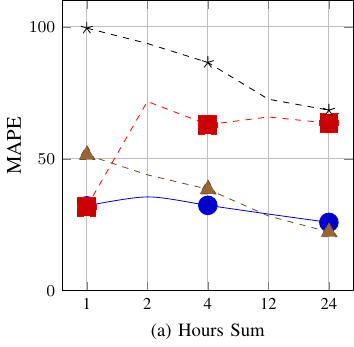}
  \caption*{}
\end{subfigure}%
\begin{subfigure}{.33\linewidth}
  \centering
  \includegraphics[width=\linewidth]{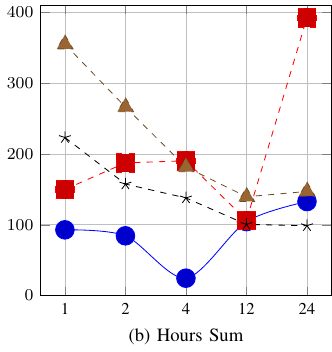}
  \caption*{}
\end{subfigure}%
\begin{subfigure}{.33\linewidth}
  \centering
  \includegraphics[width=\linewidth]{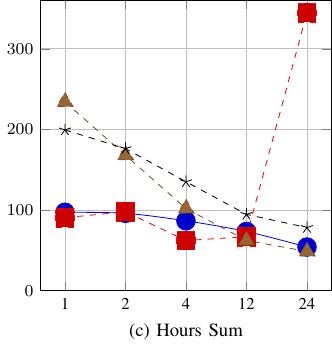}
  \caption*{}
\end{subfigure}
\\[-.1in]
\begin{subfigure}{.25\textwidth}
  \centering
  \includegraphics[scale = 0.75]{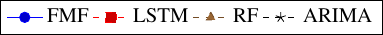}
  \caption*{}
\end{subfigure}%
\caption{\textsc{mape} comparison of  \textsc{fmf}, \textsc{lstm}, \textsc{rf}, and \textsc{arima}  with increasing hours aggregation for (a) Sweden, (b) Ireland, and (c) Australia dataset. Figure is best seen in color. } 
	\label{fig:technique_comparisons_with_hour_agg}
\end{figure}

In this section, we show results of \textsc{fmf} and other methods on forecasting loads of individual consumers for longer periods. We add consecutive hours (rows) of the original load matrix in this setting. We aggregated $2, 4, 12,$ and $24$ consecutive hours of the original load matrix (training and testing data before transformation). Figure~\ref{fig:technique_comparisons_with_hour_agg} shows the \textsc{mape} of \textsc{fmf}, \textsc{lstm}, \textsc{rf}, and \textsc{arima} with increasing hours aggregation. Clearly, with no hour aggregation (when $x$-axis value is $1$), \textsc{fmf} is better than \textsc{rf} and \textsc{arima} on all datasets. \textsc{fmf} yields better results than \textsc{lstm}  on the Ireland, comparable on the Sweden and performs slightly worse on the Australia dataset. However, performance of \textsc{lstm} degrades with increasing hours aggregation. 



\subsubsection{Forecast for clusters of households}
To evaluate the performance of \textsc{fmf} on higher spatial granularity,  we group the households into clusters and forecast the total load of each cluster for individual hours. To compare our \textsc{fmf} with~\cite{kell2018segmenting}, we follow the same train-validation-test split setting ($9$ months-$3$ months-$6$ months) for the Ireland dataset (as used in~\cite{kell2018segmenting}).  We cluster households, represented by our feature vectors (see Section~\ref{section_Matrix_Factorization}) into varying numbers of clusters ($2$ to $7$). Table~\ref{tbl:eEnergy_Comparison} shows the \textsc{mape} of \textsc{fmf} and the scheme presented in~\cite{kell2018segmenting}. \textsc{fmf} outperform  other methods in majority of the scenarios. We report improvement over \textsc{rf}, the top performing and recommended method in~\cite{kell2018segmenting}.
\begin{table}[ht!]
	\centering
	\begin{tabular}{p{1.9cm}p{0.6cm}p{0.6cm}p{0.6cm}p{0.6cm}p{0.6cm}p{0.6cm}}
		\toprule
		\multirow{2}{*}{Techniques} &\multicolumn{6}{c}{\textsc{mape} for different numbers of clusters} \\  
		\cline{2-7}
		& $2$ & $3$ & $4$ & $5$ & $6$ & $7$ \\
		\midrule 
		\textsc{nn}   & 6 & 5.3 & 5 & 5.1 & 5 & 5.2\\
		\textsc{svr}   & 6 & 5.3 & 5.1 & 5.1 & 5.2 & 5.4\\
		\textsc{rf} & 6.1  & 5 & \textbf{4.6} & 4.6 & 4.7 & \textbf{4.7}\\
		\textsc{lstm} & 10.8 & 9.2 & 9 & 8.5 & 8.7 & 8.6 \\
		\textsc{fmf}  & \textbf{4.4} & \textbf{3.2} & 4.8 & \textbf{4.3} & \textbf{4.3} & 5.6\\
				\midrule 

		\%-improvement of \textsc{fmf} over \textsc{rf}  & \multirow{3}{*}{$22.8 $}  & \multirow{3}{*}{$36$}  & \multirow{3}{*}{$-4.3 $}  & \multirow{3}{*}{$6.5$}  & \multirow{3}{*}{$8.5 $}  & \multirow{3}{*}{$-19.1$} \\
		\bottomrule
	\end{tabular}
	\caption{\textsc{mape} of \textsc{fmf} and models proposed in~\cite{kell2018segmenting} on the Ireland dataset with increasing number of household clusters.}
	\label{tbl:eEnergy_Comparison}
\end{table}

\subsubsection{Forecast of clusters of households for longer durations}
In this section, we first aggregate hours (add consecutive hours) and then households (cluster the households and aggregate their total loads) and then forecast the total load using \textsc{fmf}. We report the errors in clusters' total forecasted loads for different periods (consecutive hours sum shown on the horizontal axis) for varying numbers of clusters. We can see from Figure~\ref{fig:mape_with_hour_agg} (in appendix) that forecasting the aggregated load (both row ``hours" and column ``households" wise) helps to reduce the error in most cases. The error values show some variations, but overall the error reduces as we aggregate the hours and households. The spikes in some cases are due to randomness in loads of households. 

\section{Conclusion and Future Work} \label{Section_Conclusion}
Short-term load forecasting is a fundamental task in power system operations. \textsc{stlf} at the household level is pivotal for demand response programs such as peak shaving, dynamic pricing, and soft load shedding. In this work, we proposed a matrix factorization-based method called \textsc{fmf} for short-term load forecasting at the individual consumer level. \textsc{fmf} is computationally less expensive and can be applied to any locality because it does not use any consumers' demographic or activity pattern data. \textsc{fmf} forecast the hourly load of individual households with high accuracy compared to  state-of-the-art machine learning-based methods e.g. \textsc{lstm}, \textsc{svm}, \textsc{rf} etc. We have observed up to $49\%$ improvement in \textsc{mae}, up to $33\%$ improvement in \textsc{rmse}, up to $99\%$ improvement in \textsc{nrmse} and up to $62\%$ improvement in \textsc{mape} over other techniques for  \textsc{stlf} on different datasets. \textsc{fmf} also produce promising results on forecasting loads of clusters of consumers for longer durations. This illustrates the general ability of \textsc{fmf} to apply in any temporal and spatial granularity. Potential future work is to combine \textsc{fmf} with existing machine learning methods in an ensemble-based approach. Another future direction is to extend the approach of \textsc{fmf} to the problems of wind speed and solar intensity forecasting.

\bibliographystyle{IEEEtran}
\bibliography{load}

\pagebreak

\begin{appendices}
	\section{Implementation Details}
	A separate model is learned for each household by \textsc{arima} and \textsc{rf} while a single model is learned for all households by \textsc{lstm} due to its computational complexity. 
	\subsubsection{\textsc{arima}}
	The input parameters for \textsc{arima} are the maximum and minimum number of autoregressive terms ($p$), the maximum and the minimum number of nonseasonal differences needed for stationarity ($q$), and the maximum and minimum number of lagged forecast errors ($d$). The output is an \textsc{arima} model fitted according to {\em Akaike Information Criterion}. We select $d = 0$, minimum $p = 1$, maximum $p = 5$, minimum $q = 1$, maximum $q = 5$. We fit Seasonal \textsc{arima}  (\textsc{sarima}), which is more suitable due to capturing seasonal information, \textsc{sarima}  $(p,q,d)(P,Q,D,S)$. The value of $S$ is $24$ because of seasonality effect. Maximum $P$ and maximum $Q$ values are taken as $2$, and maximum $D$ value is taken as $0$.
	
	\subsubsection{Random Forest (\textsc{rf})}
\begin{figure}[h!]
		\centering
		\begin{subfigure}{.25\textwidth}
			\centering
			\includegraphics[width=\linewidth]{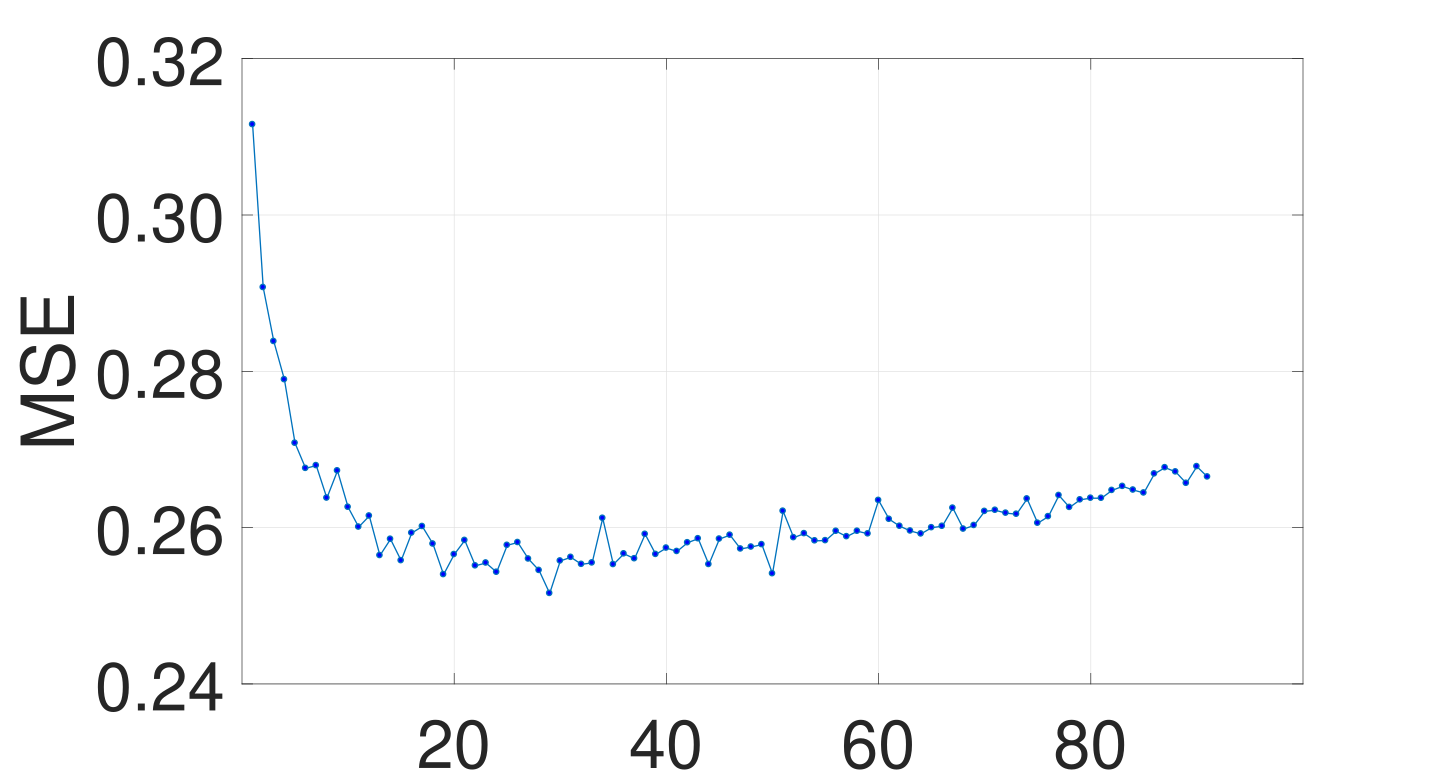}
			\caption{Attributes for Individual Tree}
		\end{subfigure}%
		\begin{subfigure}{.25\textwidth}
			\centering
			\includegraphics[width=\linewidth]{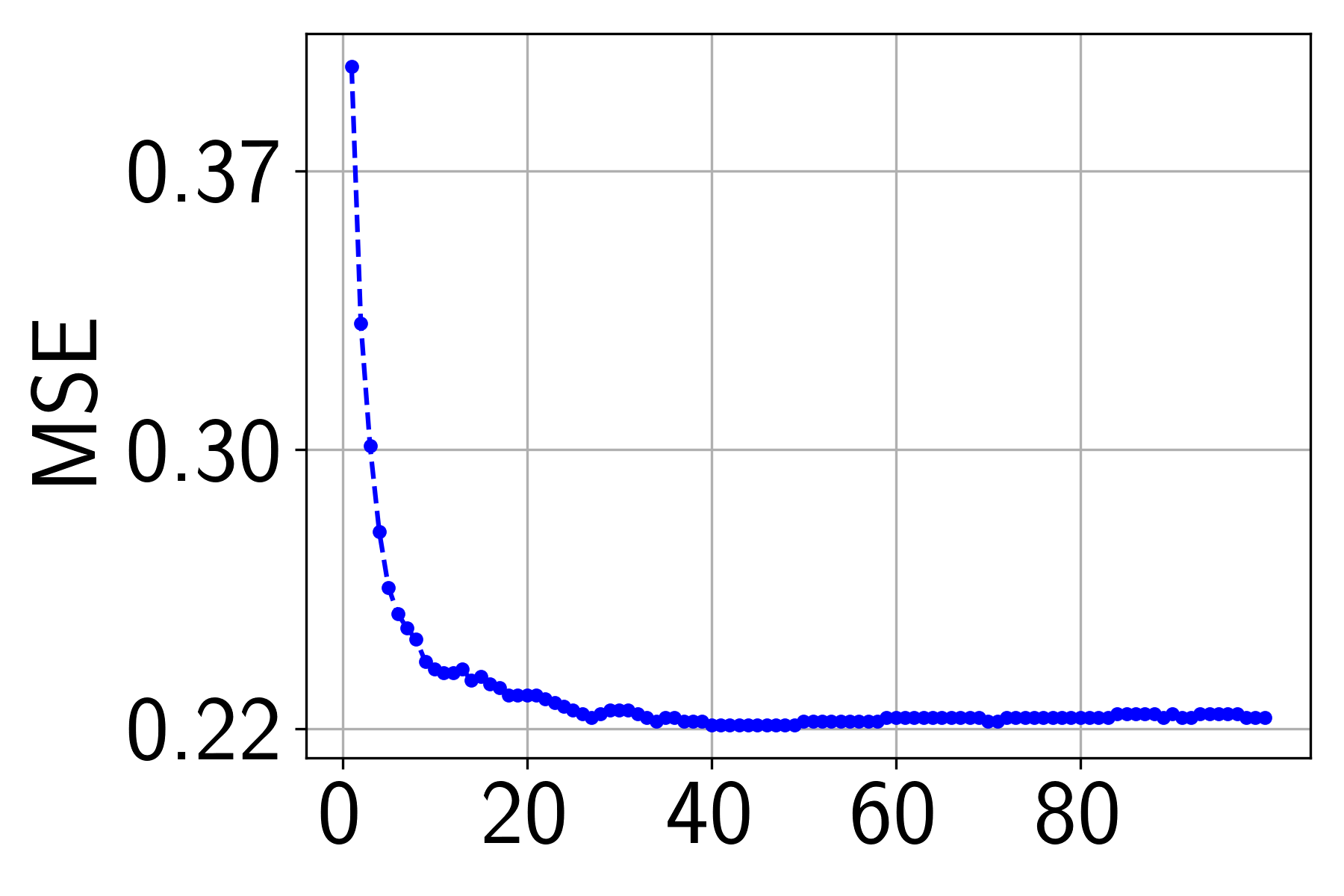}
			\caption{Total Number of Trees}
		\end{subfigure}%
		\caption{Hyperparameter tuning of Random Forest for a single Household of Sweden dataset.}
		\label{random_forest_hyper_parameter}
	\end{figure}

	Two essential hyperparameters for \textsc{rf} are 'maximum number of attributes used in a tree' and the 'the number of trees', which are tuned using a validation set. 
	As we increase the number of trees the error decreases, as shown in Figure \ref{random_forest_hyper_parameter}(b). Due to computational constraints, we select $100$ trees, and the maximum attributes used in each decision tree are $29$. This attributes count is chosen using grid search on the validation set with (\textsc{mse}) as error metric. 
	In Figure \ref{random_forest_hyper_parameter} (a), \textsc{mse} has least value against the attribute count of $29$. 
	Since \textsc{rf} is invariant to scaling, we used Table \ref{hyper_parameters_description} features without scaling. A separate model is learned (with different hyperparameters) for each consumer in each dataset.

	\subsubsection{\textsc{lstm}}
	The architecture of \textsc{lstm} scheme proposed in~\cite{kong2019short} consists of $2$ hidden layers with $20$ nodes in each layer. A simple neural network with sigmoid activation function is used to ensemble the final prediction. The features used in \textsc{lstm} are shown in Table \ref{hyper_parameters_description}. The number of previous time-stamps used in the forecasting is $6$. Furthermore, \textsc{mse} with Adam optimizer~\cite{kingma2014adam} is used to perform training for $100$ epochs. Since \textsc{lstm} contains a large number of weights, learning these weights requires huge amount of data. We test separate \textsc{lstm} models for each consumer and a single model on all consumers. Due to less amount of data, a separate model on individual consumer performs poorly compared to a single model for all consumers. Therefore, we only include the results of a single model for all consumers in each dataset. 
	\begin{table}[h!]
		\centering
		\begin{tabular}{llp{5cm}}
			\toprule
			Index &\multicolumn{1}{c}{Variable} & \multicolumn{1}{c}{Description}  \\ [0.5ex] 
			\midrule
			1-24 & Hours & One hot encoding vector\\
			25-31 & Day of Week & One hot encoding vector \\
			32-62 & Day of Month & One hot encoding vector\\
			63-74 & Month & One hot encoding vector\\
			75-85 & Lagged Input & 3 previous hours of same day, 4 hours of previous day including predicted hour predicted, 4 hours of same day previous week \\
			86 & Public Holiday & Boolean input\\
			87 & Temperature & Single numeric input\\
			88 & Wind Speed & Single numeric input \\
			89 & Humidity & Single numeric input\\
			[1ex] 
			\bottomrule
		\end{tabular}
		\caption{Input parameters for the classification algorithms.}
		\label{hyper_parameters_description}
	\end{table}

	\section{Supplementary Results}
	\begin{figure}[h!]
		\centering
		\includegraphics[width=.8\linewidth]{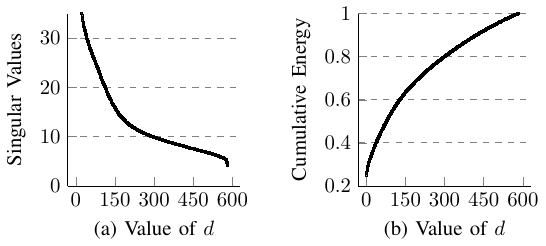}
		\caption{(a) Singular values, and (b) Cumulative energy contained in the first $d$ dimensions of $\Sigma$ (for Sweden dataset).}
		\label{fig_svd_val_swe}
	\end{figure}
	
	\begin{figure}[h!]
		\centering
		\includegraphics[width=.75\linewidth]{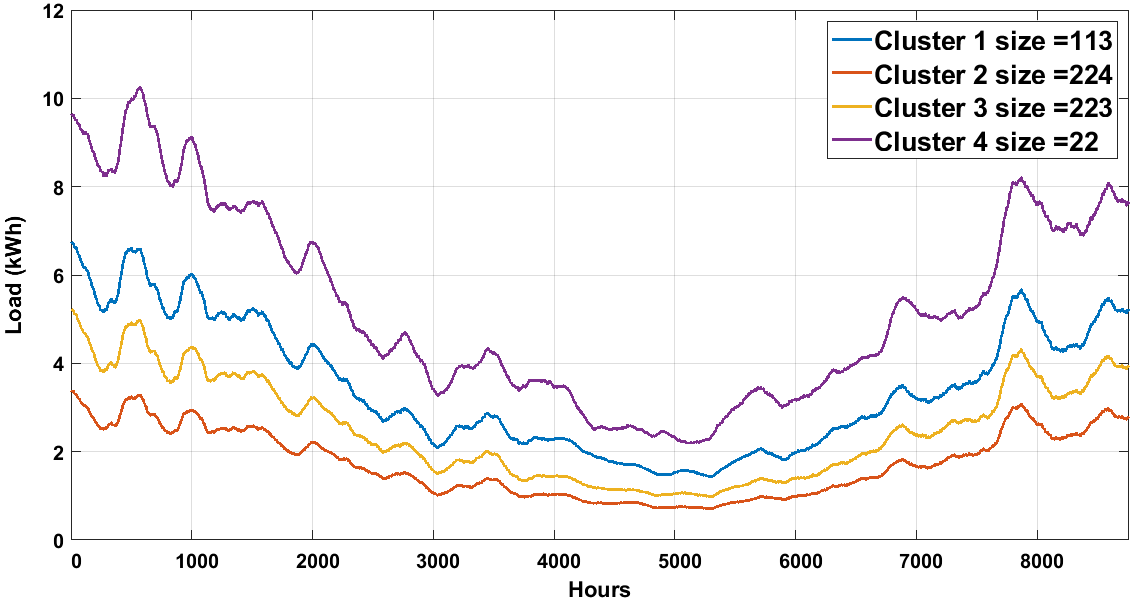}
		\caption{Total hourly loads of different clusters of consumers for one year in Sweden dataset. Figure is best seen in color.}
		\label{fig_clust_goodness_swe}
	\end{figure}
	
	\begin{figure}[h!]
		\centering
		\includegraphics[width=.65\linewidth]{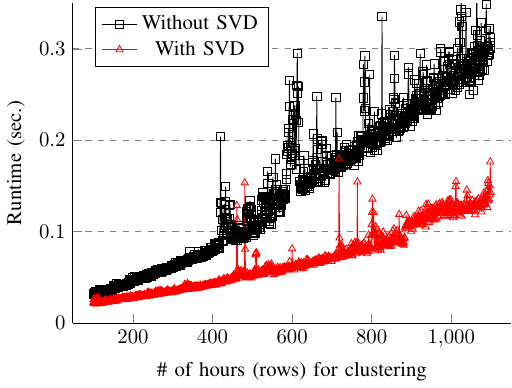}
		\caption{Runtime of $k$-means algorithm ($k=80$) on load matrix without and with dimensionality reduction. Horizontal axis shows the increasing number of hours in the training set.}
		\label{fig_clust_dim_time}
	\end{figure}
	

	
	
	\begin{figure*}[]
		\centering
		\begin{subfigure}{.245\textwidth}
			\centering
			\includegraphics[scale = 0.22] {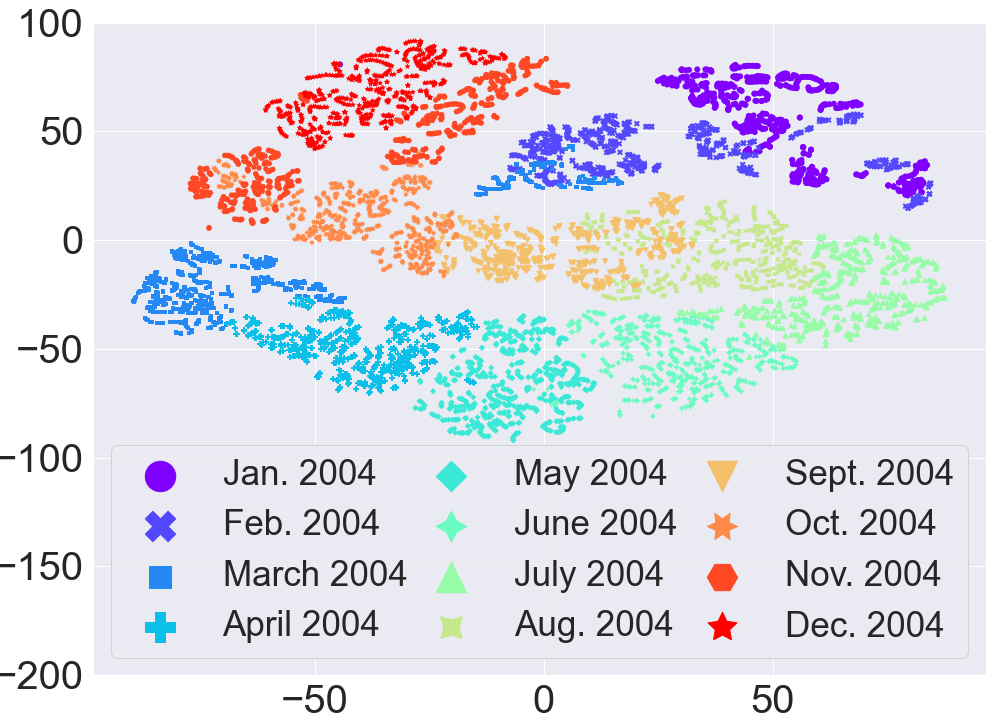}
			\caption{Month wise labels}
			\label{fig_tsne_swe_month_wise}
		\end{subfigure}%
		\begin{subfigure}{.245\textwidth}
			\centering
			\includegraphics[scale = 0.22] {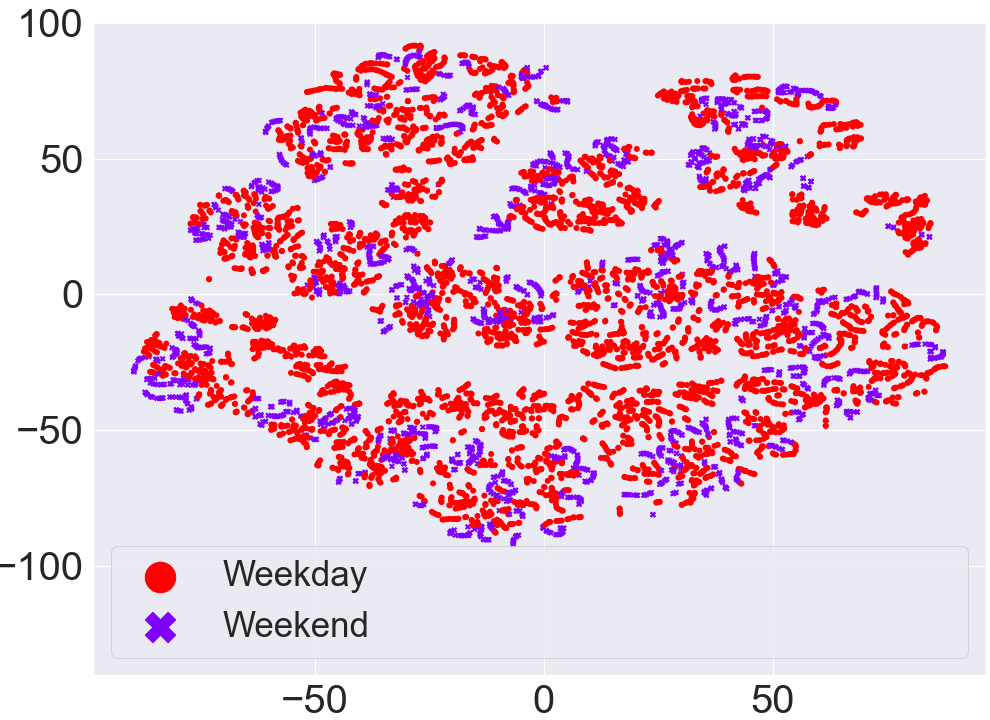}
			\caption{Weekdays/Weekends}
			\label{fig_tsne_swe_weekdays_weekends}
		\end{subfigure}%
		\begin{subfigure}{.245\textwidth}
			\centering
			\includegraphics[scale = 0.22] {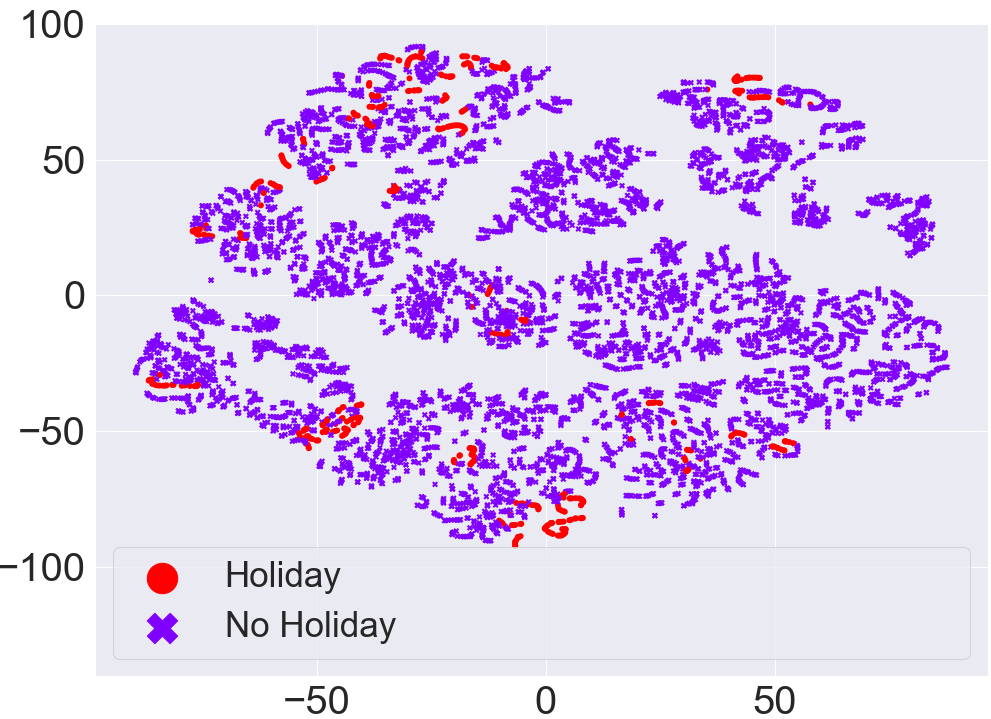}
			\caption{Public Holiday}
			\label{fig_tsne_swe_public_holiday}
		\end{subfigure}%
		\begin{subfigure}{.245\textwidth}
			\centering
			\includegraphics[scale = 0.22] {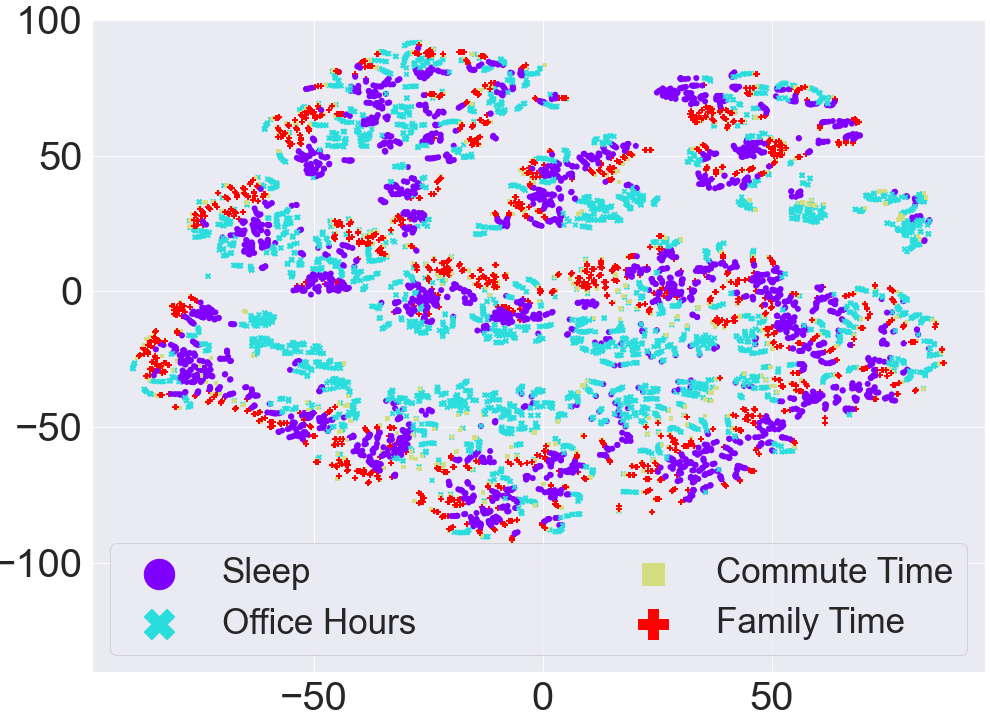}
			\caption{Hours of the Day}
			\label{fig_tsne_swe_hours}
		\end{subfigure}
		\caption{The t-\textsc{sne} plots of different labels for Sweden Dataset. Figure is best seen in color.}
		\label{fig_tsne_swe_labels}
	\end{figure*}

	\begin{figure*}[]
		\centering
		\begin{subfigure}{.245\textwidth}
			\centering
			\includegraphics[width=\linewidth] {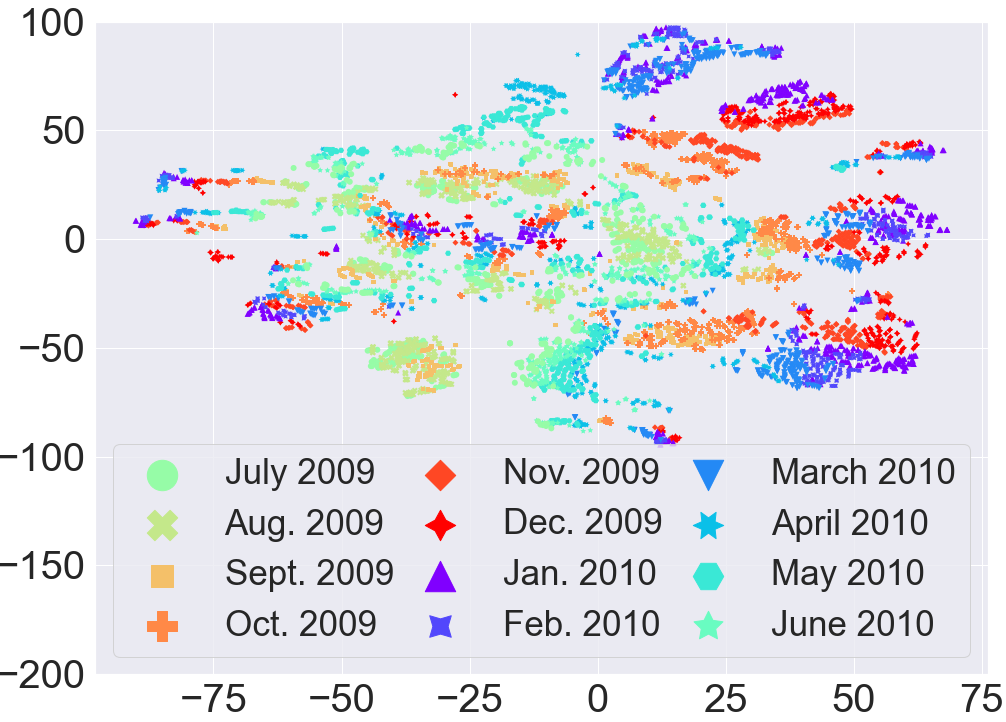}
			\caption{Month wise labels}
			\label{fig_tsne_ire_month_wise}
		\end{subfigure}%
		\begin{subfigure}{.245\textwidth}
			\centering
			\includegraphics[width=\linewidth] {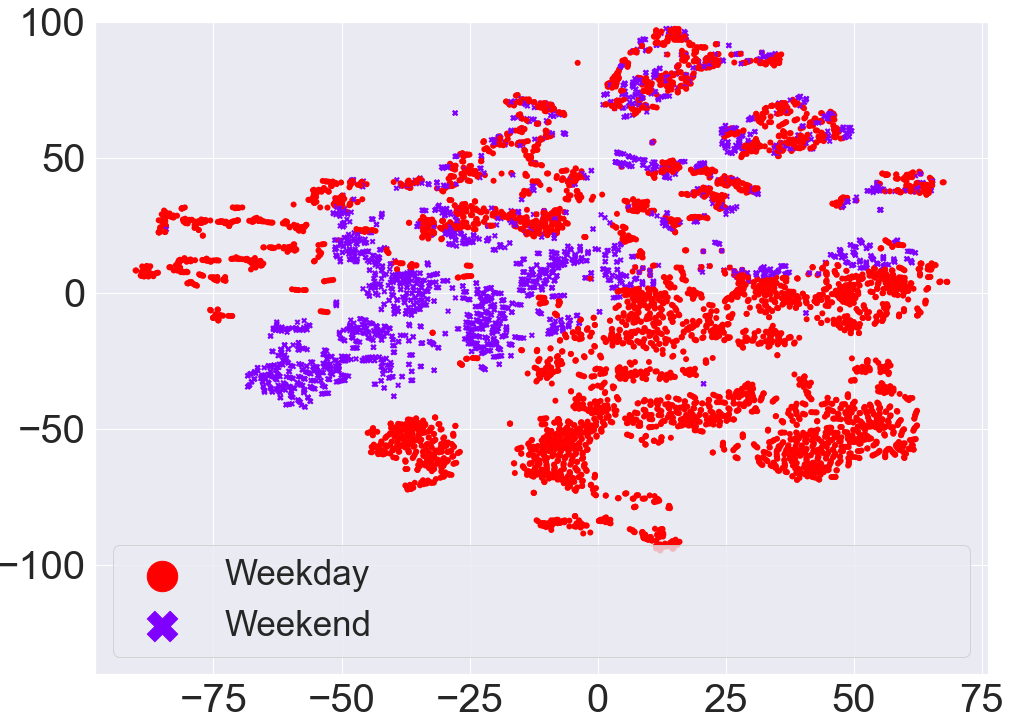}
			\caption{Weekdays/Weekends}
			\label{fig_tsne_ire_weekdays_weekends}
		\end{subfigure}%
		\begin{subfigure}{.245\textwidth}
			\centering
			\includegraphics[width=\linewidth] {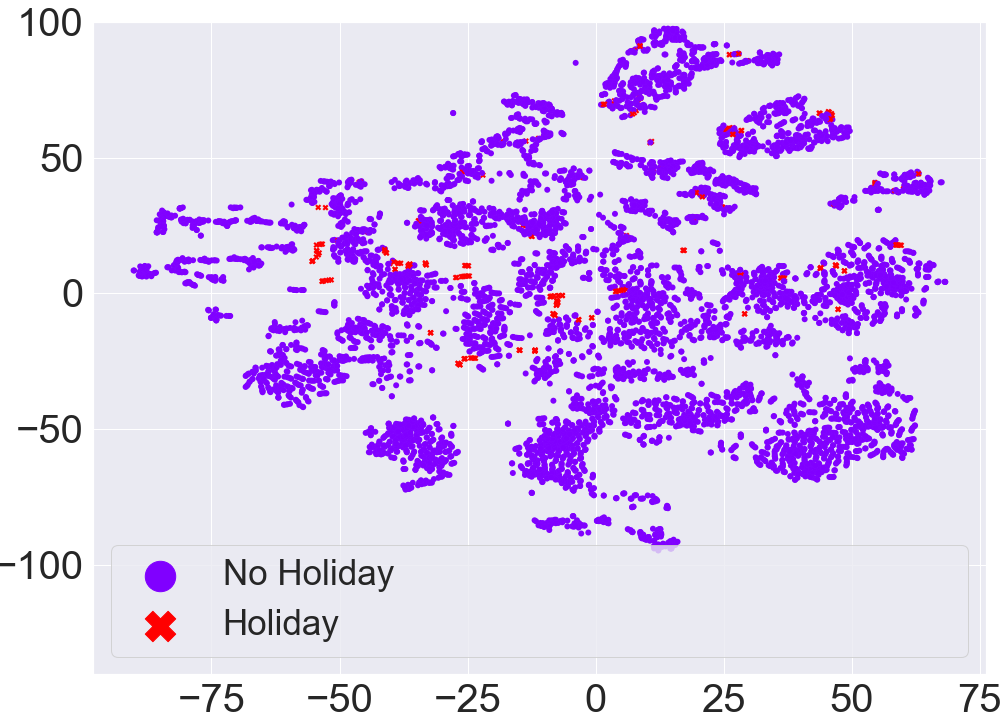}
			\caption{Public Holiday}
			\label{fig_tsne_ire_public_holiday}
		\end{subfigure}%
		\begin{subfigure}{.245\textwidth}
			\centering
			\includegraphics[width=\linewidth] {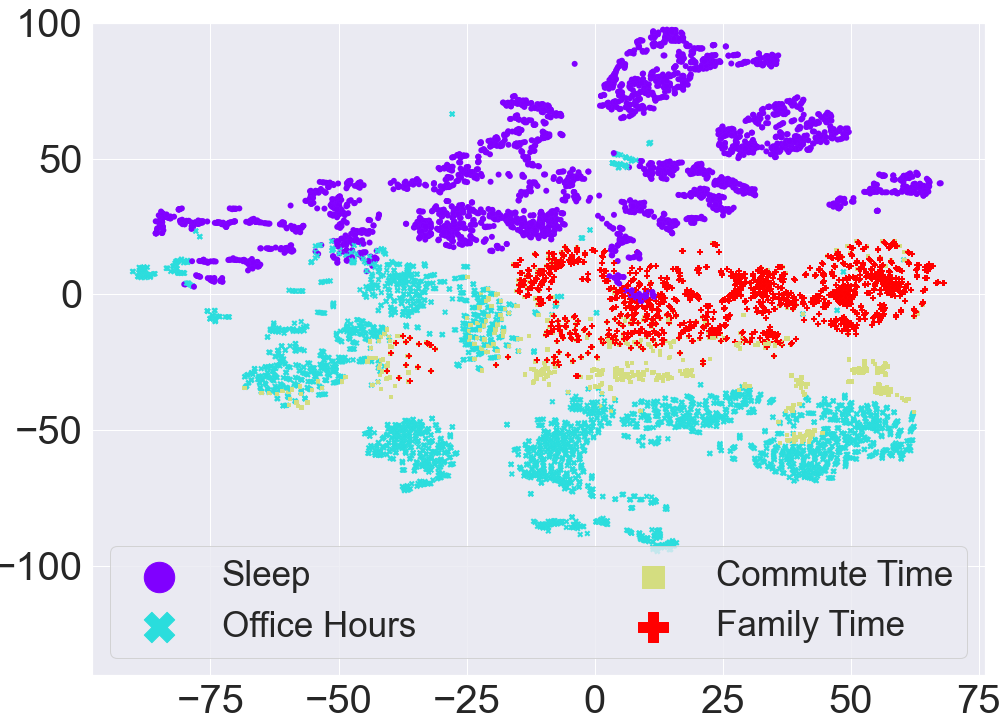}
			\caption{Hours of the Day}
			\label{fig_tsne_ire_hours}
		\end{subfigure}
		\caption{The t-\textsc{sne} plots of different labels for Ireland dataset. Figure is best seen in color.}
		\label{fig_tsne_ire_labels}
	\end{figure*}
	
	
	\begin{figure*}[]
		\centering
		\begin{subfigure}{.48\linewidth}
			\centering
			\includegraphics[width=.9\linewidth, page = 1] {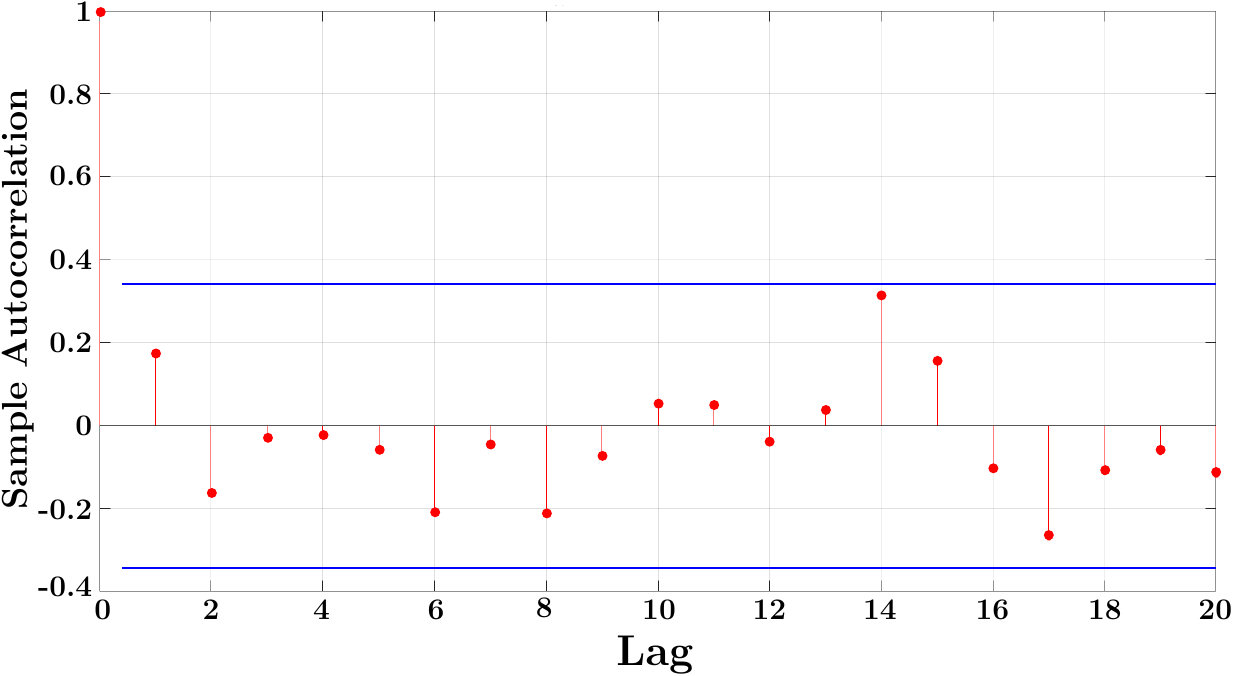}
			\caption{Australia dataset}
			\label{fig_corr_aus}%
		\end{subfigure}%
		\begin{subfigure}{.48\linewidth}
			\centering
			\includegraphics[width=.9\linewidth, page = 3] {Figures/correlation_plots.pdf}
			\caption{Sweden dataset}
			\label{fig_corr_swe}%
		\end{subfigure}
		\caption{Autocorrelation in the datasets. Sweden data has very low autocorrelation for all lag values (similarly Ireland data). The Australia data shows a slightly more significant autocorrelation.}
	\end{figure*}
	
	\begin{figure*}[h!]
		\centering
		\begin{subfigure}{.33\linewidth}
			\centering
			\includegraphics[scale=0.8]{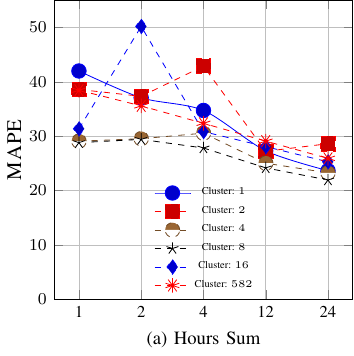}
			\caption*{}
		\end{subfigure}%
		\begin{subfigure}{.33\linewidth}
			\centering
			\includegraphics[scale=0.8]{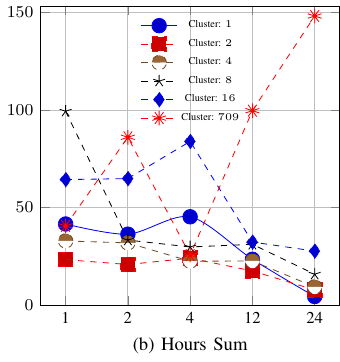}
			\caption*{}
		\end{subfigure}%
		\begin{subfigure}{.33\linewidth}
			\centering
			\includegraphics[scale=0.8]{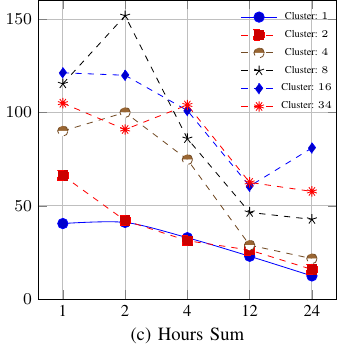}
			\caption*{}
		\end{subfigure}
		\caption{\textsc{mape} of \textsc{fmf} for groups of households and longer durations. `Cluster: $i\,$' means households are grouped into $i$ clusters. Datasets: (a) Sweden (b) Ireland (c) Australia}
		\label{fig:mape_with_hour_agg}
	\end{figure*}
	
	
	


\end{appendices}

\end{document}